\documentclass[showpacs,preprintnumbers,amsmath,amssymb]{revtex4}
\usepackage{graphicx,epsf}
\def\be{\begin{equation}}
\def\ee{\end{equation}}
\def\bea{\begin{eqnarray}}
\def\eea{\end{eqnarray}}

\def\R{{\cal{R}}}

\def\vphi{\varphi}
\def\half{\frac12}

\def\nn{\nonumber}
\begin{document}
%
\title{Cosmological perturbations through a simple bounce}
\author{Laura E.~Allen and David Wands}
%
%
%
\affiliation{
Institute of Cosmology and Gravitation, University of Portsmouth,
Portsmouth~PO1~2EG, United Kingdom}
\date{\today}
%
%
%

\begin{abstract}
We present a detailed study of a simple scalar field model that
yields non-singular cosmological solutions. We study both the
qualitative dynamics of the homogeneous and isotropic background
and the evolution of inhomogeneous linear perturbations. We
calculate the spectrum of perturbations generated on super-Hubble
scales during the collapse phase from initial vacuum fluctuations
on small scales and then evolve these numerically through the
bounce. We show there is a gauge in which perturbations remain 
well-defined and small throughout the bounce, even though 
perturbations in other commonly used gauges become large or 
ill-defined. We show that the comoving curvature perturbation
calculated during the collapse phase provides a good estimate of
the resulting large scale adiabatic perturbation in the expanding
phase while the Bardeen metric potential is dominated by what
becomes a decaying mode after the bounce. We show that a power-law
collapse phase with scale factor proportional to $(-t)^{2/3}$ can
yield a scale-invariant spectrum of adiabatic scalar perturbations in the
expanding phase, but the amplitude of tensor perturbations places
important constraints on the allowed initial conditions.
\end{abstract}

\pacs{98.80.Cq \hfill PU-ICG-04/07, astro-ph/0404441v3}

\maketitle


\section{Introduction}

\let\include\input

Current cosmological observations, especially measurements of the
spectrum of temperature anisotropies and polarisation of the cosmic
microwave background, provide strong evidence for the existence of
primordial density perturbations on large scales, far larger than the
Hubble scale at the time of last-scattering. Observations are
consistent with an almost scale-invariant, Gaussian distribution of
adiabatic primordial perturbations~\cite{WMAP}.

The standard explanation for the origin of the primordial
perturbations is that the Universe underwent an accelerated,
inflationary, expansion in the early universe~\cite{LLbook}.
Zero-point vacuum fluctuations in a massless scalar field naturally
give rise to a scale-invariant distribution of effectively classical
field fluctuations on super-Hubble scales during an exponential (de
Sitter) expansion. If inflation is driven by a light, slowly-rolling,
inflaton field then these field fluctuations can be directly related
to a comoving curvature perturbation, $\R$, which remains constant for
adiabatic perturbations on super-Hubble scales until a given
wavelength re-enters the horizon in the subsequent radiation- or
matter-dominated eras.

It is important to consider whether or not inflation is the {\em
only} consistent model for the origin of large-scale structure in
our Universe.
An alternative proposal is that the hot big bang and the
primordial perturbations originate not from an accelerated
expansion, but from a preceding collapse.
In classical general relativity a collapsing universe may be
doomed to a singularity, but in a finite quantum theory of
gravity, such as string theory, the spectrum of states going into
the ``singularity'' would be expected to determine some out-going state.
In particular the collapse phase could set the initial conditions
for a subsequent post-big bang phase. Gasperini and Veneziano
originally proposed \cite{GV} a pre big bang scenario based on the
superstring low energy effective action \cite{pbbreviews}. More
recently Khoury et al proposed the ekpyrotic scenario
\cite{ekpyrotic} motivated by colliding brane solutions in
heterotic M-theory (see also \cite{pyro,cyclic} for later
developments of this scenario). Both models invoke a collapsing
universe in the 4D Einstein frame where the energy density is
dominated by minimally coupled scalar fields.

In a collapsing universe zero-point vacuum fluctuations with a given
wavelength are squeezed outside the rapidly shrinking Hubble scale,
also naturally giving rise to a spectrum of field fluctuations on
super-Hubble scales. However in either pre big bang or ekpyrotic
scenarios, vacuum fluctuations give rise to a steeply blue-tilted
spectrum of comoving curvature perturbations during the collapse
\cite{BGGMV,Lyth}. If the tilt of the comoving curvature spectrum
remains unchanged during the transition from pre to post big bang
phase then this cannot seed the observed large scale structure in our
Universe. One would then have to consider alternative sources for
structure after a collapse phase, such as the curvaton mechanism
\cite{curvaton}, or consider a different collapse model. Collapse
driven by matter or a scalar field with pressureless equation of
state can produce a scale-invariant spectrum of comoving curvature
perturbations~\cite{Wands99,FB}.

Although very general physical arguments have been advanced that
the comoving curvature perturbation, $\R$, should be conserved on
super-horizon scales for adiabatic density perturbations
\cite{Lyth85,DM,MS,WMLL,LW03} there are several possible ambiguities that
may arise in the transition from collapse to bounce, not present
in conventional models of an inflationary expansion. At any bounce
the comoving Hubble scale diverges and all scales are inside the
Hubble scale, at least for an instant~\cite{MP}. And the Bardeen metric
potential, $\Phi$, does not have the same spectral dependence as
the comoving curvature perturbation during the collapse
\cite{BF,Hwang,Khoury2,Lyth2,HwangNoh,DV}, 
even though the two tilts must coincide for adiabatic
perturbations in the subsequent expanding universe. As a result it is
impossible for both $\R$ and $\Phi$ to obey non-singular evolution
equations through the bounce \cite{CDC}.

In fact the Bardeen potential on super-Hubble scales grows rapidly in
a collapsing phase.  This was noted in Ref.~\cite{BGGMV} who realised
that the divergence of perturbations in the Bardeen's longitudinal (or
conformal Newtonian) gauge could be eliminated by a
gauge-transformation to the spatially flat (or off-diagonal) gauge.
However we will see that at the moment of bounce the spatially flat
gauge itself becomes ill-defined. The comoving curvature perturbation
may remain small during collapse but it too becomes ill-defined in the
vicinity of the bounce in our model. The breakdown of perturbation theory
in any given gauge does not necessarily signal the breakdown of
perturbation theory. It is well known that the comoving
curvature perturbation becomes singular at stationary values of the
field even in single-field inflation models, but there are other gauge
choices where perturbations remain small~\cite{Kodama}.

Several authors have studied the evolution of perturbations through a
bounce using some analytic form for the background scale factor
\cite{PPN,PPNG,Finelli:2003mc} and/or the evolution equations for perturbations
\cite{CDC}. But this has lead to apparently contradictory results
depending on which functions are assumed to have a smooth regular form
\cite{CDC}.

In order to investigate some of these issues associated with the
evolution of cosmological perturbations through the bounce we will
consider a simple scalar field model for a non-singular bounce.  
We are following the general approach envisaged in the pre big bang
scenario \cite{GV} where higher-order corrections to the effective
action will become significant at high curvatures or strong coupling
and can yield non-singular bounce solutions \cite{nonsing,Ed}.  In the
ekpyrotic or cyclic models the transition is supposed to occur 
at a real singularity in the effective-4D theory. Not only background
quantities but also perturbations diverge here. So our 4D analysis
will not address arguments about the regularisation schemes at the
singularity proposed so far \cite{Khoury2,Schwarz,Tolley}.

There
have already been a few numerical studies of the evolution of
perturbations through non-singular bounce solutions including
higher-order corrections to the string effective action
\cite{Tsujikawa:2001ad,Tsujikawa:2002qc,CartierHwang,Cartier}.
It can be shown that that the null energy condition must be violated
for a bounce to occur in general relativity in a spatially flat 4D FRW
cosmology \cite{Brustein}.  (This is quite different from non-singular
 classical bounces which have been studied in closed FRW models
\cite{ParkerFulling,Starobinsky78,Matzner,Hwangbounce,GordonTurok,Deruelle,DeruelleStreich,MP} 
which do not violate the null energy
condition, but require fine-tuned initial conditions to obtain a
sufficiently long collapse phase \cite{Starobinsky78,Matzner}.)
%
%
One problem with results based on higher-order corrections near the
bounce is that the evolution close to the bounce becomes
dominated by the ``corrections'' and one cannot be sure that even
higher-order corrections would not also become important
\cite{Ed}. Another complication is that increasingly higher-order
corrections are equivalent to introducing additional fields with
non-canonical kinetic terms \cite{TT,Wands93} so that the true
dynamical degrees of freedom become obscured. 

In order to make the physical degrees of freedom manifest we will work
with scalar fields obeying canonical wave equations in
four-dimensional Einstein gravity. The price we must pay for such
simplicity, while still obtaining a non-singular bounce, is that one
of the scalar fields will be a massless ``ghost'' field with negative
kinetic energy, in order to violate the null energy condition near the
bounce. Such a ghost field would
lead to the presence of serious instabilities in any low energy theory
\cite{ghosts}. However it does not play any significant role in our model at
early or late times. Instead it leads to precisely the sort of
instability at high energies that we require to achieve a bounce in
the high-energy regime, destabilising the pre big bang collapse phase
and triggering the transition to an expanding cosmology. 
Our scalar field model is intended only as a simple model of a
non-singular bounce and not as a realistic model of the low energy universe.
The early- and late-time (low energy) evolution in our model is
dominated by a conventional minimally-coupled scalar field with
exponential potential. By choosing the dimensionless slope of the
potential we are able to investigate different power-law solutions for
the pre big bang collapse leading to different power spectra for the
field and metric perturbations \cite{Wands99}. A very similar bounce
model was studied by Peter and Pinto-Neto \cite{PPN}, but we find very
different results.

In the original pre big bang scenario higher-order corrections to the
effective action will become significant at high curvatures or strong
coupling and can yield non-singular bounce solutions \cite{nonsing,Ed},
at least for some initial conditions and some higher-order
corrections. In particular it can be shown that the null energy
condition must be violated for a bounce to occur in a 4D FRW cosmology
\cite{Brustein}.


We first study the background evolution of our model using a
phase-plane analysis in Section \ref{background}. With a particular
choice of dimensionless phase variables the early or late time
solutions correspond to critical points in the phase plane. However an
alternative choice of variables is better suited to study evolution
through the bounce. We identify a class of non-singular models which
start in a power-law collapse in the asymptotic past, pass smoothly
through a bounce and approach a power-law expansion.

We then define linear perturbations about this background model in
Section \ref{perturb}. We present the evolution and constraint
equations for the perturbations in arbitrary gauge and then identify
particular gauge choices that are well-suited to study the
perturbations at early and late times, or through the bounce,
analogously to the different choice of background variables used to study
the background phase-plane in different regimes. The choice of gauge
to take the perturbations through the bounce is vital as it must be
ensured that the gauge itself is well defined throughout the bounce,
and we identify problems with gauges commonly used in inflationary
models. 

In section \ref{evolve} we evolve first-order evolution equations for
the metric and field perturbations and use the constraint equations as
a consistency check of our calculation. We assume perturbations begin
in the quantum vacuum state on small scales and use this to calculate
the spectrum of perturbations generated on super-Hubble scales during
collapse. We evolve this numerically through the bounce in order to
calculate the resulting spectrum of curvature perturbations. We
present our results in section \ref{results} in terms of the
gauge-invariant variables ${\cal R}$ and $\Phi$. We find that it is
possible to produce a scale-invariant spectrum of adiabatic
perturbations on super-Hubble scales from a collapse phase in our
simple model. Finally we present our conclusions in Section
\ref{conc}.

\section{Background}
\label{background}

We consider a spatially flat FRW cosmology containing two
minimally scalar fields; one with positive kinetic energy and
self-interaction potential and one with negative kinetic energy
and zero potential energy.
The Lagrangian of the system is
 \be
\mathcal{L} = - \half\varphi^{,\mu}\varphi_{,\mu} - V(\varphi) +
\half\chi^{,\mu}\chi_{,\mu} \,.
 \ee
With our choice of signature the field $\varphi$ has a positive definite
energy density and $\chi$ is a ``ghost'' field with negative energy density.

We take the potential of the field to be a simple exponential
 \be
V=V_0\exp(-\lambda\kappa\varphi)\,. \label{exppot}
 \ee
Varying the action with respect to each of the fields and the
metric then gives a Klein-Gordon evolution equation for each of
the fields and an acceleration equation
 \bea
\varphi''+2h{\varphi'}+a^2V_{,\varphi}&=&0 \label{kgphi}\,,\\
{\chi''}+2h{\chi'}&=&0 \label{kgchi}\,,\\
h'&=&\frac{\kappa^2}{3}(-{\vphi'}^2+{\chi'}^2+a^2V) \label{evol} \,.
 \eea
subject to the Friedmann constraint
 \be
h^2 = \frac{\kappa^2}{3}\left(\half{\varphi'}^2 -\half{\chi'}^2+a^2V
\right)\label{fried} \,,
 \ee
where a prime denotes differentiation with respect to conformal
time, $\eta$.

We can integrate Eq.~(\ref{kgchi}) to obtain the
first integral
 \be
 \label{defK} \chi' = \frac{K}{a^2}\,.
 \ee
Thus the negative energy density of the ghost field decreases as
the universe expands and grows as it contracts, inversely
proportional to the square of the volume,
$\chi^{\prime2}/a^2\propto a^{-6}$.

The overall equation of state of the two scalar field system is
 \bea \label{defw}
w&=&\frac{P}{\rho}\nn\,,\\
&=&\frac{\half{\varphi'}^2-\half{\chi'}^2-a^2V}{\half{\varphi'}^2-\half{\chi'}^2+a^2V}
\,, \eea which can be written as
 \be
 w = -1 +
\frac{3\kappa^2(\varphi^{\prime2}-\chi^{\prime2})}{3h^2} \,.
 \ee
Thus we have a "phantom" equation of state ($w<-1$)
\cite{Caldwell} whenever the total kinetic energy of the fields
goes negative, i.e., for $\varphi^{\prime2}<\chi^{\prime2}$.

\subsection{Asymptotic phase-space variables}

To describe the asymptotic behaviour of the system we define the
dimensionless phase-space variables \cite{CEW}
\bea
x&\equiv&\frac{\kappa{\varphi'}}{\sqrt{6}h}\,, \nn\\
y&\equiv&\frac{\kappa a\sqrt{V}}{\sqrt{3}h}\,,\label{defxyz}\\
z&\equiv&\frac{\kappa{\chi'}}{\sqrt{6}h} \nn\,.
\eea

The Friedmann equation (\ref{fried}) then gives a constraint equation
 for the system
 \be
x^2+y^2-z^2=1 \,.
 \ee
This reduces the system to become two dimensional, with all the
solutions lying on the two-dimensional surface of a unit
hyperboloid in three-dimensional Euclidean phase space.

The Klein-Gordon equations (\ref{kgphi})
and~(\ref{kgchi}) for the two fields and the evolution
equation~(\ref{evol}) give the first-order evolution
equations for the phase variables
 \bea
 \label{dxdN}
\frac{dx}{dN}&=&-3x(1-x^2+z^2)+\lambda \sqrt{\frac{3}{2}}y^2\,,\\
\frac{dy}{dN}&=&y(3x^2-3z^2-\lambda \sqrt{\frac{3}{2}}x )  \,, \\
 \label{dzdN}
\frac{dz}{dN}&=&-3z(1-x^2+z^2) \,.
 \eea
where $N\equiv \ln a$ is the number of e-folds of expansion of the
universe.

\begin{figure}
\begin{center}
\includegraphics[width=70mm, height=80mm]{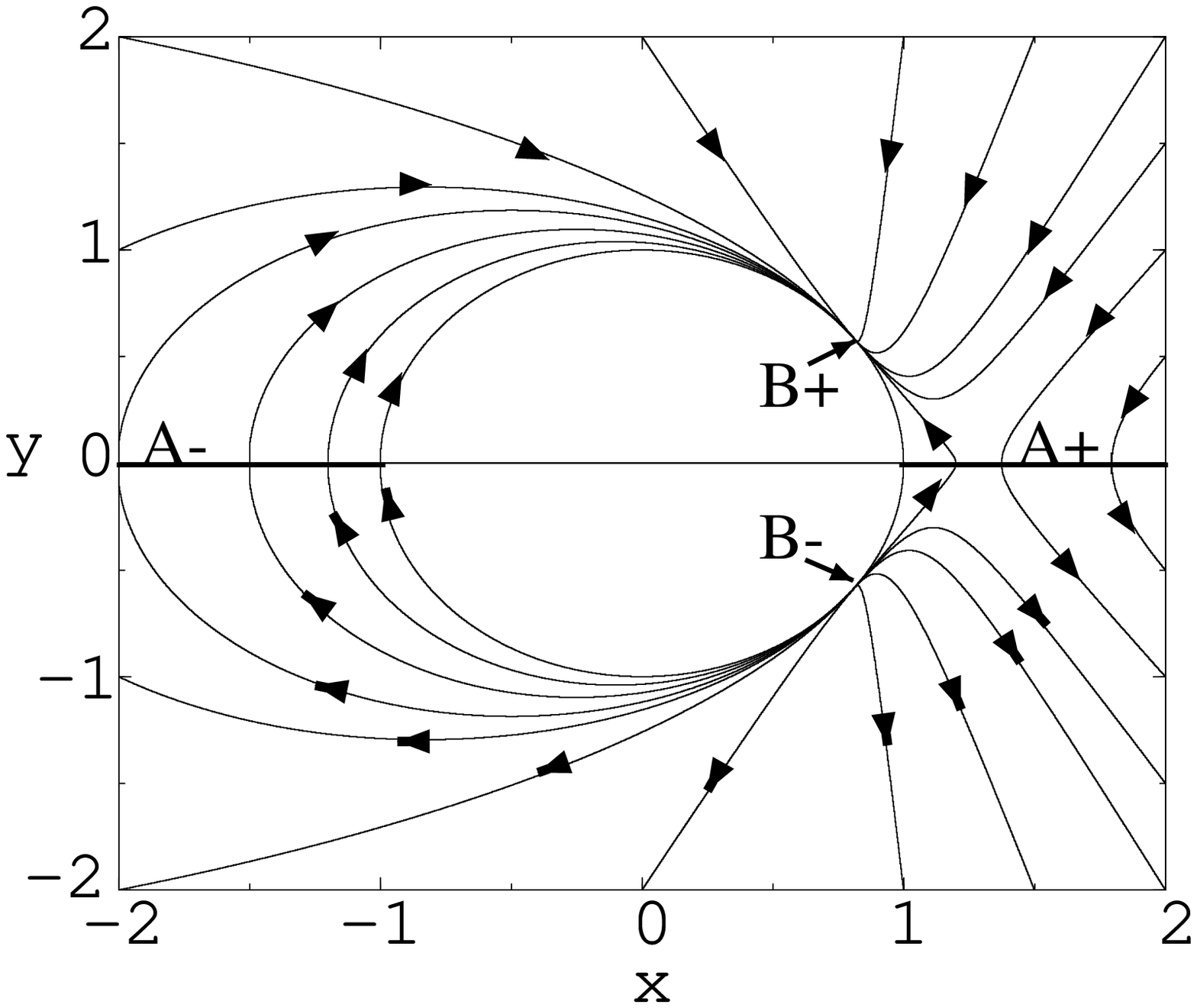}
\includegraphics[width=70mm, height=80mm]{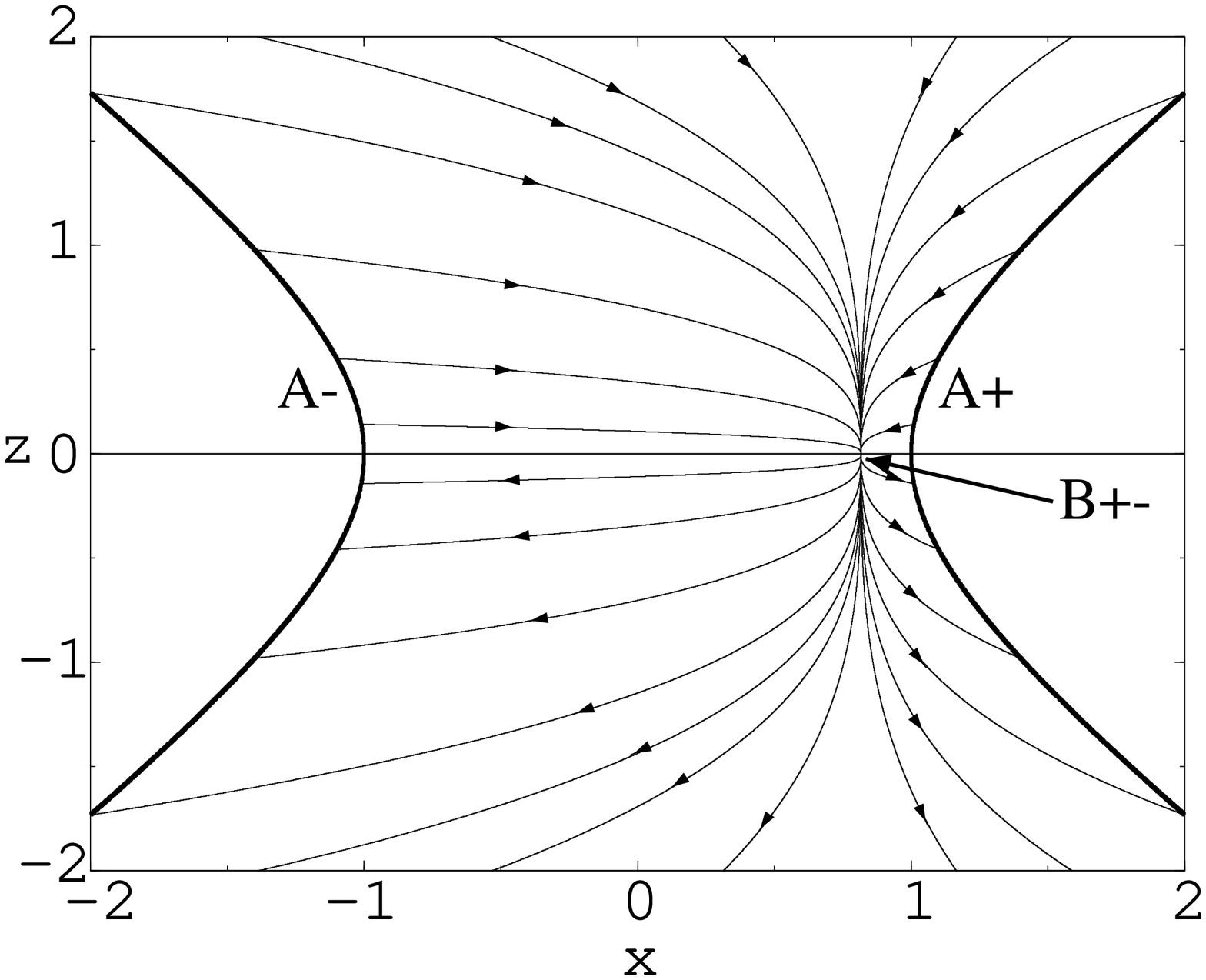}
\caption[xyzplane]{\label{xyzplane} The phase trajectories in the
(x,y) and (x,z) phase-planes for $\lambda=2$.}
\end{center}
\end{figure}

This system of equations can be solved numerically to find the
phase-trajectories. In figure \ref{xyzplane} we show phase
trajectories for $\lambda=2$ projected onto the y-x and z-x
planes. Without loss of generality we take $\chi'>0$, as from
Eq.~(\ref{defK}) $\chi$ must either be monotonically increasing or
decreasing.

The fixed points of the system (critical points where
$\frac{dx}{dN}=\frac{dy}{dN}=\frac{dz}{dN}=0$) are at
 \bea
A_{\pm}:~ &x=x,~~y=0,~~z=\pm\sqrt{x^2-1} \,,&
 \\
 \label{pointB}
B_{\pm}:~ &x=\frac{\lambda}{\sqrt{6}},~~y=\pm\sqrt{1-\frac{\lambda^2}{6}},~~z=0
\,.
 \eea
following the notation of \cite{Heard}.

$A_{\pm}$ are hyperbolic lines of critical points ($|x|\geq1$)
where the potential energy $V(\varphi)$ is negligible. The
universe is dominated by the kinetic energy of the two fields with
$\varphi'\propto\chi'\propto h$. The scale factor evolves as a
power-law with respect to conformal time, $a\propto |\eta|^{1/2}$
(or $a\propto |t|^{1/3}$ in terms of proper time, $dt=ad\eta$).

Critical points $B_{\pm}$ describe a power-law solution where the
density of the phantom field $\chi$ is negligible, and the
potential and kinetic energy of the $\varphi$ field are
proportional. This is the well-known scaling solution for
exponential potentials\cite{LM85,BB,Halliwell}
\begin{equation}
\label{powerlaw}
 a\propto |t|^p \,, \qquad p=2/\lambda^2 \,.
\end{equation}
Note that for a positive potential $V_0>0$ (which we assume here)
$B_{\pm}$ only exist for $\lambda<\sqrt{6}$. Thus, the phase-plane
evolution will fall into two distinct parameter regimes defined by
$\lambda\lessgtr\sqrt{6}$. Within either of these regimes however,
the precise value of $\lambda$ will not affect the qualitative
description of the evolution.
Particular cases of interest are $\lambda=2$ for which the
universe described by point $B_{\pm}$ evolves like a
radiation-dominated universe with $P=\rho/3$, or
$\lambda=\sqrt{3}$ for which the universe evolves like a
dust-dominated universe with $P=0$.

We find that whenever $B_\pm$ exits they are a late-time attractor
as $N\to\infty$, i.e., at late times in an expanding universe or
early times in a contracting universe.

We will be interested in the non-zero set of collapsing universes
which start at point $B_-$ in the asymptotic past. This solution
is unstable as we approach $a\to0$ and generic solutions either
collapse to a kinetic-dominated singularity at $A_\pm$ or bounce,
at which point $|y|\to\infty$. There is another non-zero set of
collapsing solutions which start at $A_+$ before bouncing, but we
shall focus on models where the energy of the ghost field is
negligible, $z\to0$, at both early and late times.

The phase-plane (\ref{dxdN}--\ref{dzdN}) gives a useful representation
of the evolution at finite $h$ and in particular clearly identify the
asymptotic fixed points at early and late times. However the variables
(\ref{defxyz}) are liable to diverge at any bounce point where $h=0$.
Thus to study the evolution through the bounce we seek an alternative
choice of variables such the variables remain finite at the bounce.

\subsection{Phase-space variables near bounce}\label{altphase}

Close to the bounce we will describe the expansion $h$, and kinetic
energy and potential of the $\varphi$ field relative to the kinetic
energy of the $\chi$ field. {}From Eq.~(\ref{defK}) we can see that
the kinetic energy of the $\chi$ field remains non-zero and is
greatest at the bounce.
We can rearrange the Friedmann equation to get
\be
\label{friedrearranged}
1=\frac{{\varphi'}^2}{{\chi'}^2}+\frac{2a^2V}{{\chi'}^2}-
\frac{6h^2}{\kappa^2{\chi'}^2} \,,
\ee
and define phase variables
\bea
\alpha&=& \frac{{\varphi'}}{{\chi'}} \label{defalpha} \,, \\
\beta&=&\frac{a\sqrt{2V}}{{\chi'}}   \,, \\
\gamma&=&\frac{\sqrt{6}h}{\kappa{\chi'}} \label{defgamma}\,.
\eea
which are all non-singular through the bounce.
Then Eq.~(\ref{friedrearranged}) gives the constraint
\be
\alpha^2+\beta^2-\gamma^2=1 \,,
\label{abcconstraint}
\ee
so all the phase trajectories lie on the two-dimensional surface of a
unit hyperboloid. The bounce ($\gamma=0$) corresponds to the neck of the
hyperboloid ($\alpha^2+\beta^2=1$).

At the same time it is useful to change the ``time'' variable to a new
variable $q$ defined by
\be
q\equiv \frac{\kappa \chi}{\sqrt{6}} \,,
 \label{defq}
\ee
so q is a dimensionless representation of the scalar field $\chi$. The
advantage of this is that, unlike $N$, $q$ will be either
monotonically increasing or decreasing through the bounce and for all
time, as can be seen from Eq.~(\ref{defK}).

\begin{figure}
\begin{center}
\includegraphics[width=70mm]{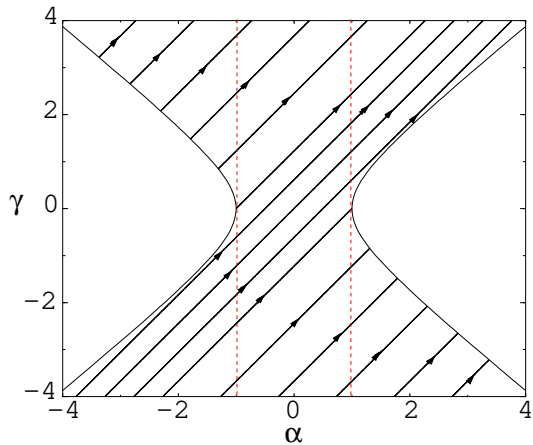}
\caption[alphagamma]{\label{alphagamma}
Phase-space trajectories in the $(\alpha,\gamma)$ plane for $\lambda=2$. 
Everywhere within the red dashed lines the universe will have a phantom 
equation of state.
}
\end{center}
\end{figure}

The system of equations (\ref{kgphi}--\ref{evol}) then becomes
\bea
\frac{d\alpha}{dq}&=&\frac{\lambda\sqrt{6}}{2}\beta^2\label{dalphadq}\,,\\
\frac{d\beta}{dq}&=&\beta(3\gamma-\frac{\lambda\sqrt{6}}{2}\alpha)
\label{dbetadq}\,,\\
\frac{d\gamma}{dq}&=&3\beta^2\label{dgammadq} \,.
\eea
Equations (\ref{dalphadq}) and~(\ref{dgammadq}) can be integrated to
give a first integral
\be
 \label{defC}
\alpha - \frac{\lambda}{\sqrt{6}}\gamma = C \,,
\ee
where $C$ is a constant. Thus the solutions evolve along straight
lines in the $(\alpha,\gamma)$ plane, as shown in Figure~(\ref{alphagamma}).

The only fixed points of this system are at $\beta=0$, i.e., they
lie on the hyperbolae
\be
A_{\pm}:~~\alpha=\pm\sqrt{1+\gamma^2} \,,
\ee
which describes the kinetic-dominated scaling solution,
$\vphi'\propto\chi'\propto h$.

As before, the phase plane evolution falls into two categories
depending on whether $\lambda\lessgtr\sqrt{6}$. In terms of the
variables $\alpha$ and $\gamma$ we see that this corresponds to the
slope of the trajectories being greater or less that unity. This
determines whether or not trajectories hit the hyperbolae $A_\pm$.

\begin{figure}
\begin{center}
\includegraphics[width=70mm]{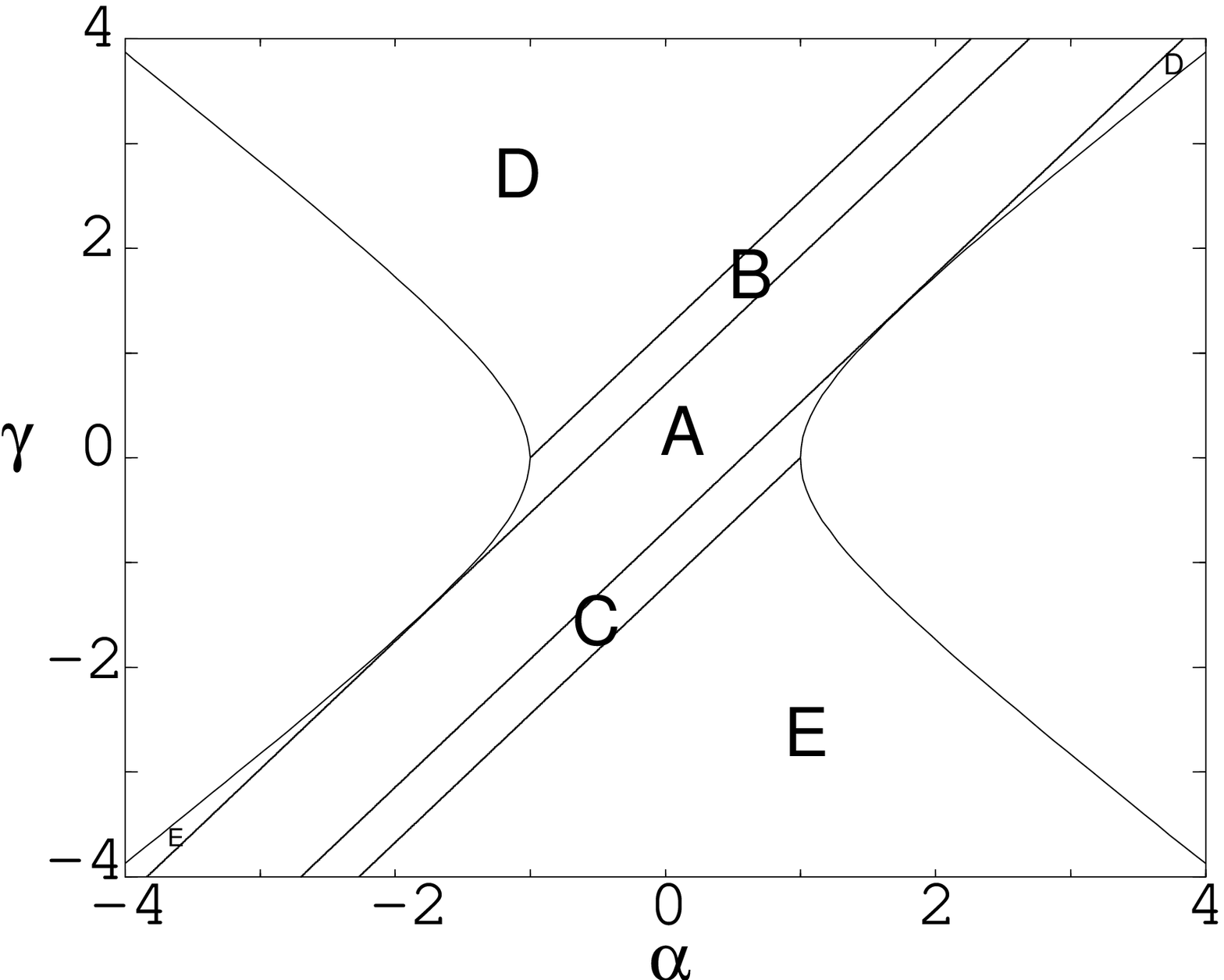}
\caption[alphagamma]{\label{acschematic}
Classification of phase-space trajectories in the $(\alpha,\gamma)$
plane for $\lambda=2$.}
\end{center}
\end{figure}

For $\lambda^2<6$ the phase plane looks like that shown in
Figure~\ref{alphagamma}. The trajectories fall into different classes
according to the choice of initial conditions. This is shown
schematically in Figure~\ref{acschematic}.

The evolution of the universe in each of the classes is
\begin{itemize}
\item[{\textsf A}]: power-law contraction ($p=2/\lambda^2$) $~\rightarrow~$ bounce
$~\rightarrow~$ power-law expansion ($p=2/\lambda^2$)
\item[{\textsf B}]: kinetic dominated contraction $~\rightarrow~$ bounce
 $~\rightarrow~$ power-law expansion ($p=2/\lambda^2$)
\item[{\textsf C}]: power-law contraction ($p=2/\lambda^2$) $~\rightarrow~$ bounce
$~\rightarrow~$ kinetic dominated expansion
\item[{\textsf D}]: Big Bang $~\rightarrow~$ kinetic dominated expansion
$~\rightarrow~$ power-law expansion ($p=2/\lambda^2$)
\item[{\textsf E}]: power-law contraction ($p=2/\lambda^2$) $~\rightarrow~$ kinetic
dominated contraction $~\rightarrow~$ Big Crunch
\end{itemize}

\begin{figure}
\begin{center}
\includegraphics[width=70mm]{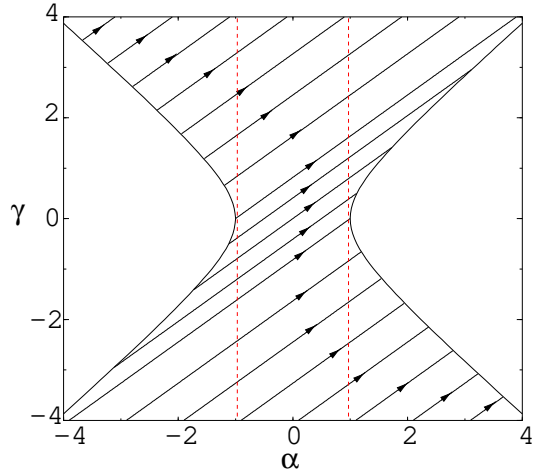}
\caption[aclambda3]{\label{aclambda3}
Phase-space trajectories in the $(\alpha,\gamma)$ plane for $\lambda=3$. 
Everywhere within the red dashed lines the universe will have a phantom 
equation of state.
}
\end{center}
\end{figure}

\begin{figure}
\begin{center}
\includegraphics[width=70mm]{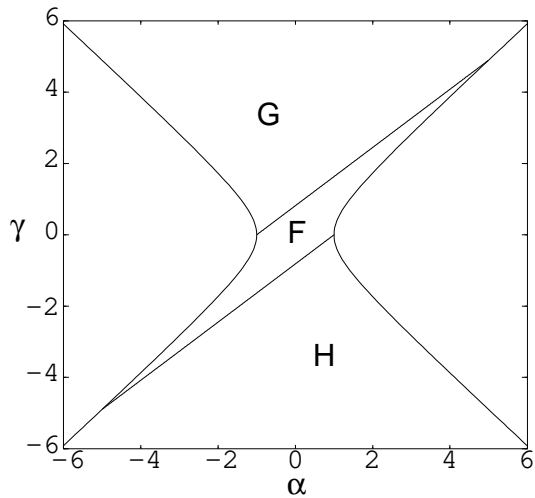}
\caption[aclambda3schem]{\label{ac3schematic}
Classification of phase-space trajectories in the $(\alpha,\gamma)$
plane for $\lambda=3$}
\end{center}
\end{figure}

For $\lambda^2>6$ the phase trajectories evolve as shown in
Figure~\ref{aclambda3}.  The different classes of solution are shown
schematically in Figure~\ref{ac3schematic}.
The trajectories fall into only three classes in this case. The evolution
of the universe in each of the classes is:
\begin{itemize}
\item[{\textsf F}]: kinetic dominated contraction $~\rightarrow~$ bounce
$~\rightarrow~$ kinetic dominated expansion
\item[{\textsf G}]: Big Bang $~\rightarrow~$ kinetic dominated expansion
\item[{\textsf H}]: kinetic dominated contraction $~\rightarrow~$ Big Crunch
 \end{itemize}


The overall equation of state (\ref{defw}) of the system is
\be
w=-1 + \frac{2(\alpha^2-1)}{\gamma^2}\,.
\ee
In order to have phantom behaviour ($w<-1$) we therefore require
$\alpha^2<1$.
For $\lambda=0$, the trajectories are $\alpha=const$, and so are
straight vertical lines in the $\gamma-\alpha$ phase plane. For
$\lambda\neq0$ no trajectory remains within the phantom region and we
always have $w>-1$ at early times and late times.

In the second part of the paper we will study inhomogeneous
perturbations about the solutions in class A, i.e., non-singular
evolution for $\lambda^2<6$ where the ghost field is negligible at
early and late times. From Eqs.~(\ref{abcconstraint}) and (\ref{defC}) we find that 
this corresponds to solutions with integration constant
\be
C^2<1-\frac{\lambda^2}{6}\,.
\label{Climit}
\ee


\section{Linear perturbations}
\label{perturb}

The most general metric of a perturbed FRW universe can be written
to first order
\be
\label{gmunu}
g_{\mu\nu}=a^{2}(\eta) \left(
\begin{array}{cc}
-(1+2\phi) & B_{|i}-S_i \nonumber\\
B_{|j}-S_j &
 (1-2\psi)\gamma_{ij} + 2E_{|ij} +F_{i|j} + F_{j|i} + h_{ij}
 \nonumber
\end{array}
\right)\, .
\ee
The scalar quantity $\phi$ describes the perturbation in the lapse function,
$\psi$ describes the perturbation of the intrinsic 3-curvature and
the scalar shear is given by
\be
\sigma = E'-B \,.
\ee
We need to consider only the scalar and tensor parts ($h_{ij}$) of the
metric perturbations, which remain decoupled to first-order. Vector
perturbations ($S_i$ and $F_i$) can be eliminated by a gauge
transformation when the energy-momentum tensor only comes from scalar
fields~\cite{MFB92}. 

We can also express the perturbed fields in terms of their homogeneous
background value plus an inhomogeneous perturbation
\bea
\varphi&=&\varphi_0 +\delta\varphi\,,\\
\chi&=&\chi_0+\delta\chi \,.
\eea

Under a first-order time-shift, or temporal gauge transformation,
\be
\tilde\eta=\eta +\xi^0 \,,
\ee
the scalar metric perturbations transform as \cite{MFB92}
\bea
\tilde{\phi}&=&\phi-h\xi^0-{\xi^0}' \label{phishift}\,,\\
\tilde{\psi}&=&\psi +h\xi^0 \label{psishift}\,,\\
\tilde{\sigma} &=& \sigma + \xi^0 \,,
\eea
and the first-order field perturbations transform as
\bea
\tilde{\delta\varphi} &=& \delta\varphi - \varphi'\xi^0 \,,\\
\tilde{\delta\chi} &=& \delta\chi - \chi'\xi^0  \label{chishift} \,.
\eea
The tensor metric perturbations, $h_{ij}$, are gauge-invariant.

The linearly perturbed Klein-Gordon equations for the two fields give
the wave equations for the field perturbations:
\bea
\delta\vphi'' +2h\delta\vphi' -\nabla^2\delta\vphi
+a^2V_{,\vphi\vphi}\delta\vphi +
2a^2V_{,\vphi}\phi-\vphi_0'\phi'-\vphi_0'(3\psi'-\nabla^2\sigma) &=&
0\label{pertkgphi}\,,\\
\delta\chi''
+2h\delta\chi'-\nabla^2\delta\chi-\chi_0'\phi'-\chi_0'(3\psi'-\nabla^2\sigma)&=&0\label{pertkgchi}\,.
\eea
Similarly the Einstein equations can be split into background and
perturbed parts, yielding the background Eqs.~(\ref{evol})
and~(\ref{fried}), and to first-order
\bea
-3h(h\phi+\psi')+\nabla^2[\psi+h\sigma] &=&
 -\frac{\kappa^2}{2}(\phi({\vphi_0'}^2-{\chi_0'}^2)-
\vphi_0'\delta\vphi'+\chi_0'\delta\chi'-a^2V_{,\vphi}\delta\vphi)
 \label{rel1} \,,\\
h\phi+\psi'&=&-\frac{\kappa^2}{2}(\chi_0'\delta\chi-\vphi_0'\delta\vphi)
 \label{rel2} \,,\\
(2h'+h^2)\phi+h\phi'+\psi''+2h\psi' &=&
\frac{\kappa^2}{2}\left[\right.\vphi_0'\delta\vphi'-\chi_0'\delta\chi'-
\phi({\vphi_0'}^2-{\chi_0'}^2)-a^2V_{,\vphi}\delta\vphi\left.\right]
 \label{rel3} \,,\\
\sigma' + 2h\sigma + \psi-\phi &=&0
 \label{rel4} \,.
\eea
An arbitrary inhomogeneous perturbation can then be decomposed into a
superposition of independent Fourier modes with comoving wavenumber
$k$, such that $\nabla^2\delta\varphi=-k^2\delta\varphi$.

The gauge-transformation equations (\ref{phishift}--\ref{chishift})
show that the evolution equations contain an unphysical gauge mode,
identified with the choice of constant-time hypersurfaces, as well as
the physical degrees of freedom. To remove this ambiguity one can work
with gauge-invariant variables, but there is no unique choice of
gauge-invariant variables, such as, for example, the curvature
perturbation.

To determine the initial vacuum fluctuations at early times in a
collapsing phase and then follow these perturbations through the
bounce phase we will solve the perturbations in two specific choices
of gauge, each adapted to one phase of the evolution. Either gauge can
be used to define different gauge-invariant variables \cite{MW1}.
Because we have gauge-invariant definitions in each phase we can
consistently follow the perturbations through from early to late
times.

Ultimately we will present our final results in terms of the two most
commonly used gauge-invariant curvature perturbations:
\bea
\R &=& \psi + \frac{h(\varphi'\delta\vphi-\chi'\delta\chi)}{
  \varphi^{\prime2} - \chi^{\prime2}}
 \label{defR} \,,\\
\Phi&=&\psi+h\sigma\,.\label{defPhi}
\eea
Equation~(\ref{defR}) is the natural generalisation of the usual
comoving curvature perturbation for two scalar fields
\cite{GBW,Gordon} to the case where one of them is a ghost field,
while Eq.~(\ref{defPhi}) is the curvature perturbation in the
longitudinal gauge~\cite{MFB92}, also called the Bardeen
potential.

\subsection{Uniform curvature gauge}

In order to describe the early time behaviour close to the fixed point
$B_-$ we will first work in the uniform curvature (or spatially flat)
gauge in which the metric perturbation, $\psi$ vanishes. This is the
gauge commonly used to describe vacuum fluctuations generated during
inflation~\cite{Sasaki,LythStewart,hwang}.

{}From Eq.~(\ref{psishift}) we see that $\tilde\psi=0$ corresponds to
a specific time-shift from an arbitrary gauge
 \be
\xi^0 = - \frac{\psi}{h}
 \label{unipsishift} \,,
\ee
This enables us to give a gauge-invariant definitions of the remaining
metric perturbations \cite{MW1} and, in particular, the scalar field
perturbations in the uniform-curvature gauge
\bea
\tilde{\delta\varphi}|_\psi &\equiv& \delta\varphi + \frac{\varphi'}{h} \psi
\,,\\
\tilde{\delta\chi}|_\psi &\equiv& \delta\chi + \frac{\chi'}{h} \psi
\,.
\eea
In the rest of this sub-section we will drop the tilde and
$\psi$-subscript, but it should be remembered that all perturbations
correspond to those in the uniform-curvature gauge.

Using Eqs.(\ref{rel1}--\ref{rel4}) to eliminate $\phi$ and $\psi$, the
perturbed Klein-Gordon equations in the uniform-curvature gauge can be
written as two coupled second-order equations for the field
perturbations:
\bea
 {\delta\vphi''} -\nabla^2 {\delta\vphi}+2h {\delta\vphi'}+
\left[a^2V_{,\vphi\vphi}+\kappa^2\vphi_0'\left(\vphi_0'(\frac{h'}{h^2}-2)-
2\frac{\vphi_0''}{h}\right)\right] {\delta\vphi}&=&
\frac{\kappa^2\chi_0'}{h}(\vphi_0'\frac{h'}{h}-\vphi_0'') {\delta\chi}
\label{uc1} \,,\\
 {\delta\chi''} -\nabla^2 {\delta\chi}+2h {\delta\chi'}-
\kappa^2\chi_0'^2(2+\frac{h'}{h^2}) {\delta\chi}&=&
\frac{\kappa^2\chi_0'}{h}(\vphi_0''-\frac{h'}{h}\vphi_0') {\delta\vphi}
\label{uc2}\,.
\eea

These evolution equations in the uniform-curvature gauge have
a simple form which is convenient for studying the evolution during
monotonic expansion or collapse. In particular we will use this gauge
to describe the initial quantum fluctuations during a collapse phase.

However these equations are {\em not} suitable for evolving the
perturbations through a bounce. The evolution equations~(\ref{uc1})
and (\ref{uc2}) contain singular coefficients as $h\to0$. This is not
surprising as from Eq.~(\ref{unipsishift}) we see that the
uniform-curvature gauge is itself ill-defined at a bounce as $h\to0$.
Instead, we have to find another gauge which remains well-defined at
the bounce in order to evolve the initial vacuum perturbations through
to an expanding phase.

\subsection{Uniform $\chi$-field gauge} \label{uniformchi}

Analogously to the background solutions where we used the $\chi$-field
as a useful time coordinate to study evolution through the bounce, we
will use the uniform-$\chi$ gauge to follow linear perturbations
through the bounce.

{}From Eq.~(\ref{chishift}) we see that to set $\tilde{\delta\chi}=0$
we must choose
\be
 \label{xichi}
 \xi^0= \frac{\delta\chi}{\chi'} \,.
\ee
The remaining $\varphi$-field perturbation in the uniform-$\chi$ gauge
has the gauge invariant definition
\be
\tilde{\delta\varphi}|_\chi \equiv \delta\varphi -
\frac{\varphi'}{\chi'} \delta\chi \,.
\ee
The scalar metric perturbations $\phi$, $\psi$ and $\sigma$ in the
uniform-$\chi$ gauge also have gauge-invariant definitions. In
particular the curvature perturbation on uniform-$\chi$ hypersurfaces
is given by
\be
\tilde{\psi}|_\chi \equiv \psi + \frac{h}{\chi'}\delta\chi \,.
\ee
In the rest of this sub-section we will drop the tilde and
$\chi$-subscript, but it should be remembered that all perturbations
correspond to those in the uniform-$\chi$ gauge.

The perturbed Klein-Gordon and Einstein equations
(\ref{pertkgphi}--\ref{rel4}), then give us the four evolution
equations for the four perturbation variables
$\delta\vphi,~\psi,~\phi$ and $\sigma$ in the uniform $\chi$-field
gauge
 \bea
\label{vphiqq}
\kappa\delta\vphi_{,qq}+\frac{6k^2}{\kappa^2{\chi'}^2}\kappa\delta\vphi
+3\beta^2\lambda^2\kappa\delta\vphi-6\lambda\beta^2\phi &=&0
 \,,\\
\psi_{,qq}-\frac{3\alpha}{\sqrt6}\kappa\delta\vphi_{,q}-
\frac{3\lambda\beta^2}{2}\kappa\delta\vphi
+\gamma\phi_{,q}+3\beta^2\phi&=&0
 \,,\\
\label{phiq}
\phi_{,q}+3\psi_{,q}+\frac{6 k^2}{\kappa^2{\chi'}^2}s&=&0
 \,,\\
\label{sq}
 s_{,q}+4\gamma s+\psi-\phi&=&0\,, \eea
plus two constraint equations
\bea
\gamma\phi+\psi_{,q}-\frac{\sqrt6 \alpha}{2}\kappa\delta\vphi&=&0
\label{con1}\,,\\
(\alpha^2-1)\phi -
\frac{\alpha}{\sqrt6}\kappa\delta\vphi_{,q} +
(\frac{\lambda\beta^2}{2} - \frac{3\alpha\gamma}{\sqrt6})
\kappa\delta\vphi - \frac{2k^2}{\kappa^2{\chi'}^2} (\psi+\gamma s)
 &=& 0 \label{con2} \,.
\eea
Here we have (as for the background evolution) we have written time
derivatives with respect to the dimensionless, and monotonic, variable
$q$ defined in Eq.~(\ref{defq}).
We have written background quantities in terms of the dimensionless
variables $\alpha$, $\beta$ and $\gamma$ defined in
Eqs.~(\ref{defalpha}--\ref{defgamma}).

We have rescaled the shear perturbation variable $\sigma$ in terms of
\be
s \equiv \frac{h}{\gamma} \sigma \,,
\ee
so that all gradient terms enter through terms of order
$(k/\kappa\chi')^2$ which is the square of $\gamma$ times the ratio of
the Hubble length to the wavelength. Thus we can obtain solutions
order-by-order in terms of $(k/\kappa\chi')^2$ to define a
long-wavelength limit. Although the Hubble length diverges at the
bounce we have from Eq.~(\ref{defK}) that $(k/\kappa\chi')^2\propto
a^4$. This remains finite and actually reaches a minimum at the
bounce. Thus we find that the long-wavelength solution, neglecting
terms of order $(k/\kappa\chi')^2$ is an excellent solution close to
the bounce for scales of interest.

In principle we could use the constraint equations (\ref{con1})
and~(\ref{con2}) to eliminate $\phi$ and $s$ respectively and leave
two second-order equations for $\delta\varphi$ and $\psi$, similar to
what was done for the uniform curvature gauge perturbations in the
preceding subsection. However as $\gamma\to0$ approaching a bounce
this approach would be very sensitive to any small numerical error.
We will not eliminate $\phi$ and $s$ and instead will use their
first-order evolution equations to evolve them numerically through the
bounce. The constraint equations (\ref{con1}) and~(\ref{con2}) then
provide a useful consistency check for our numerical integrations. A 
similar argument in favour of using coupled first-order equations to
study perturbations through a bounce was recently given by 
Cartier~\cite{Cartier}.

\subsection{Tensor perturbations}

The tensor part of the perturbed metric, $h_{ij}$, is transverse and
trace-free and is invariant under gauge transformations. We expand
$h_{ij}$ in plane-waves 
 \be 
h_{ij}({\bf x},\eta) = (2\pi)^{-3}\int d^3{\bf k} \sum_{p=\times,+} \
 \delta g^{p}_{\bf k}(\eta) \,
 \epsilon^{p}_{ij}({\bf k,x}) \,, 
 \ee 
where $\epsilon^{p}_{ij}$ is the polarisation tensor and
$p=\times,~+$ represents the two independent polarisation states.
$\delta g^{\times,+}_{\bf{k}}$ is the scalar amplitude of each state
for wave-mode $\bf{k}$.
For both polarisation states this scalar amplitude obeys the simple
wave equation for a free scalar field in an FRW spacetime,
\be
\delta g''+2h\delta g' +k^2\delta g=0\,,
\label{deltageta}
\ee
where $k^2={\bf k.k}$. Here and from now on the subscript $k$ and the
superscript $\times,+$ are assumed for $\delta g$.

Under transformation to derivatives with respect to our monotonic time
variable $q$ defined in Eq.~(\ref{defq}), the tensor evolution
Eq.~(\ref{deltageta}) becomes
\be
\delta g_{,qq}+\frac{6k^2}{\kappa^2{\chi'}^2}\delta g=0\,.
 \label{deltagq}
\ee
This is the form of the equation that we use to evolve the tensor
perturbations numerically through the bounce.

\section{Vacuum fluctuations}
\label{evolve}

\subsection{Early-time behaviour in uniform-curvature gauge}\label{earlyuc}

For any power-law solution described by a critical point in the
dimensionless phase-space we have $\varphi' \propto h$ and the
right-hand-sides of Eqs.~(\ref{uc1}) and~(\ref{uc2}) vanish.  Thus
the field perturbations $\delta\varphi$ and $\delta\chi$ in the
uniform-curvature gauge decouple in the early-time limit where the
background solution is described by Eq.~(\ref{powerlaw}).
In this case the evolution equations for the field perturbations
can be written very simply in terms of the canonical
Sasaki/Mukhanov variables \cite{MFB92}
 \bea
u_\varphi &=& a\delta\varphi \,,\\
u_\chi &=& a\delta\chi \,,
 \eea
For each Fourier mode with comoving wavenumber $k$, 
Eqs.~(\ref{uc1}) and~(\ref{uc2}) can then be written as
 \be
u_i''+\left(k^2-\frac{a''}{a}\right)u_i=0 \,,
\label{uweqn}
 \ee
where $i=\varphi,\chi$.
The evolution equation for the tensor perturbations
Eq.~(\ref{deltageta}) can be written in exactly the same form, where
\be
u_g = \frac{a\delta g}{2 \kappa} \,.
\ee

For power law expansion or collapse described by
Eq.~(\ref{powerlaw})
the equation (\ref{uweqn}) 
has the general solution
\cite{LythStewart}
 \be
u
 =
 \sqrt{|k\eta|}[u_{+}H_{(|\nu|)}^{(1)}(|k\eta|)+u_{-}H_{(|\nu|)}^{(2)}(|k\eta|)] 
 \label{uwsoln} \,,
 \ee
where the index
 \be
\label{defnu}
\nu \equiv \frac{3}{2}+\frac{1}{p-1} \,.
 \ee
 Although the scalar fields $\varphi$ and $\chi$ are real scalar
 fields, we have chosen to Fourier expand the perturbations in terms
 of complex Fourier modes $\propto e^{\pm ikx}$. Thus we work with the
 complex modes functions (\ref{uwsoln}) which must obey the condition
 $u(-k)=u^*(k)$.

At early times ($\eta\ll-k^{-1}$) we have free oscillations
$u\propto e^{\pm ik\eta}$. The simplest assumption is that the
perturbation variables, $u_{1,2}$, start in the flat space-time
vacuum state with only positive frequency modes, satisfying
the normalisation condition
 \be
u'u^{*}-{u^{*}}'u=i \,.
 \ee
Thus at early times ($k\eta\to-\infty$) we require
 \be
u \to \frac{1}{(2k)^{\frac{1}{2}}}e^{-ik\eta} \label{vaclimit} \,.
 \ee

Using the vacuum solution (\ref{vaclimit}) at early times to set
the initial conditions for the general solution (\ref{uwsoln})
yields
\be
u =
 \frac{\sqrt\pi}{2} \frac{e^{i(|\nu|+\half)\frac{\pi}{2}}}{k^{\half}}
 |k\eta|^\half H^{(1)}_{|\nu|}(|k\eta|) \,. \label{vacsoln}
 \ee

We use this analytic vacuum solution for the three perturbation
variables to set up the initial conditions for the perturbations
in our numerical code. We start the code far away from the bounce
when we are still well within the power-law regime
($p\simeq2/\lambda^2$). However, as we saw in Section
\ref{background}, this power-law solution is unstable and as the
universe collapses the evolution will eventually diverge away from
the critical point.

\subsubsection{Initial power spectra}

$\delta\vphi$, $\delta\chi$ and $\delta g$ are independent
Gaussian random fields whose distribution will be completely
determined by the power spectra, defined for a scalar $\delta x$ as\footnote
{Because of the two polarisation states, the tensor power spectrum
  ${\cal P}_g\equiv k^3|\delta g|^2/\pi^2$.}
 \be
\mathcal{P}_x\equiv \frac{k^3 |\delta x|^2}{2\pi^2}
 \label{powerdef}\,.
 \ee
The tilt of the spectrum is defined by
 \be
\Delta n_x\equiv \frac{d\ln(\mathcal{P}_x)}{d\ln(k)}\,,
 \ee 
so that $\Delta n_x=0$ for a scale-invariant spectrum. Note that the
conventional spectral index for scalar metric perturbations is
given by $n=1+\Delta n_x$.

As $\delta\vphi$ and $\delta\chi$ are independent fields in the
initial power-law regime, we can find the power spectrum for any
quantity by evolving separately different modes corresponding to an
initial perturbation in $\varphi$ or $\chi$. The power due to each mode can be
summed to obtain the final spectrum. $\varphi$-modes correspond to
initial vacuum perturbations of the $\vphi$-field on the uniform
curvature hypersurface with the $\chi$ perturbation (and its
derivative) initially zero on this hypersurface. $\chi$-modes
correspond to the opposite situation, where there is an initial
perturbation of the $\chi$-field but no perturbation of the
$\vphi$ field (or its derivative) on the uniform curvature
hypersurface.

Any complex variable can be split into two parts whose phase
differs by 90$^\circ$, which typically might be real and imaginary
parts.  We find it convenient to split the Hankel function, $H_{|\nu|}^{(1)}$, 
found in our general solution for $u$, Eq.~(\ref{vacsoln}), into real Bessel 
functions of the first and second kind:
\begin{equation}
H_{|\nu|}^{(1)}(|k\eta|) = J_{|\nu|}(|k\eta|)+iY_{|\nu|}(|k\eta|) \,.
\end{equation} 
We will thus refer to these as the $J$ and $Y$ modes.
{}From the definition of the power spectrum
Eq.~(\ref{powerdef}) it can then be seen that
 \be
\mathcal{P}_x=\mathcal{P}_{x_J}+\mathcal{P}_{x_Y}
 \ee
 Thus, the $J$ and $Y$ parts can be evolved separately to
 determine the total power spectrum. This gives us four independent
 'modes' or sets of initial conditions ($\varphi_J$, $\varphi_Y$,
 $\chi_J$, and $\chi_Y$) for which we present our numerical results.

The Bessel functions of first and second kind describe regular and
singular solutions on large scales ($|k\eta|\rightarrow 0$) behaving as
\bea
J_{|\nu|}(|k\eta|) &\rightarrow&
\frac{1}{\Gamma(|\nu|+1)}\left(\frac{|k\eta|}{2}\right)^{|\nu|} \,,
\\
Y_{|\nu|}(|k\eta|) &\rightarrow&
 -
\frac{\Gamma(|\nu|)}{\pi}\left(\frac{|k\eta|}{2}\right)^{-|\nu|}
 \,.
 \eea
Thus, from Eqs~(\ref{vacsoln}), we see that the power spectrum
from vacuum fluctuations during collapse on large scales yields
 \be
\mathcal{P}_{u}
 = C_J^2(|\nu|)k^2|k\eta|^{1+2|\nu|}
  + C_Y^2(|\nu|)k^2|k\eta|^{1-2|\nu|}
\label{poweru}\,,
 \ee
where
 \bea
C_J(|\nu|)&\equiv& \frac{1}{2^{|\nu|+\frac{3}{2}}\sqrt\pi\Gamma(|\nu|+1)} \,,\\
C_Y(|\nu|)&\equiv& \frac{2^{|\nu|-\frac{3}{2}}\Gamma(|\nu|)}{\pi^{\frac{3}{2}}}\,.
 \eea
The spectral tilts for each term in the spectrum (\ref{poweru})
are then \cite{LythStewart,Wands99}
 \bea
\Delta n_{u_J}&=& 3+2|\nu| \,,\\
\Delta n_{u_Y}&=& 3-2|\nu| \,.
 \eea
 
In a contracting universe the second $Y$-term in Eq.~(\ref{poweru})
should quickly dominate on super-Hubble scales as $k\eta\to0$.
However, we will evolve both modes numerically to see whether this will
remains true through the bounce or whether the less dominant mode in
the collapsing phase will come to dominate in the expanding phase.

We will present detailed numerical solutions for interesting physical
choices of the scalar field slope, $\lambda$, and hence power-law $p$
in the collapse phase that give simple analytic solutions for the
perturbations during the initial power-law collapse phase.

\subsubsection{$\lambda=2$: Power-law solution with $p=1/2$}

Taking $\lambda=2$ means that the scale factor will follow the
same evolution as a radiation dominated universe with $w=1/3$ and
$p=1/2$ at the critical points $B_\pm$.

The Hankel function index in Eq.~(\ref{defnu}) is $\nu=-1/2$ and
for this special case the first Hankel function has the simple
form
\be
H^{(1)}_{\half}(z)=\sqrt{\frac{2}{\pi z}}(\sin(z)-i\cos(z)) \,.
 \ee
and the vacuum solutions (\ref{vacsoln}) become
 \be
u = \frac{1}{(2k)^\half}(\cos(|k\eta|)+i\sin(|k\eta|))
 \ee
In this case the $Y$-part of the solution corresponds to the real
part of the solution which will dominate in the collapsing universe
as $k\eta\to0$.

The spectral tilts of the $Y$ and $J$ parts of $u$ are then given by
 \bea
\Delta n_{u_Y}&=& 2 \,,\\
\Delta n_{u_J}&=& 4 \,.
 \eea
Both modes have steep blue tilts.

\subsubsection{$\lambda=\sqrt3$: Power-law solution with $p=2/3$}

Taking $\lambda=\sqrt3$ means that the scale factor will follow
the same evolution as a matter dominated universe at the critical
points $B_\pm$, with $w=0$ and $p=2/3$.

The Hankel function index in Eq.~(\ref{defnu}) is $\nu=-3/2$ and
for this special case the first Hankel function has the simple
form
 \be H^{(1)}_{3/2}(z) = -\sqrt{\frac{2}{\pi
z}}\left(1-\frac{1}{iz}\right)(\cos(z)+i\sin(z)) \,.
 \ee
and the solution for $u$ will be
 \be
u=\frac{1}{(2k)^\half}
\left[ \left(\cos(|k\eta|)-\frac{\sin(|k\eta|)}{|k\eta|}\right)
 + i\left(\sin(|k\eta|)+\frac{\cos(|k\eta|)}{|k\eta|}\right)  
 \right]
 \ee
This is the same mode function as is found for a
vacuum fluctuations of a massless field in de Sitter inflation
\cite{Wands99}.
Here the $Y$-mode is the imaginary part of the solution will dominate in
a collapsing universe as $k\eta\to0$.

The spectral tilts of the $Y$ and $J$ parts of the field
fluctuations on large scale are
 \bea
\Delta n_{u_Y}&=& 0 \,,\\
\Delta n_{u_J}&=& 2 \,.
 \eea
The $J$ part has a steep blue tilt, but the $Y$ part, which
dominates as $k\eta\to0$, has a scale-invariant
spectrum~\cite{Starobinsky,Wands99,FB}.


\subsection{Numerical evolution through the bounce}

The initial conditions for the background variables are set so
that the evolution follows a trajectory in the phase-plane that
lies within region A, as defined in section(\ref{altphase}). Thus
the evolution starts close to the critical point $B_-$ where it
follows a power-law collapse solution. It eventually diverges away
from this critical point as the universe collapses and the
(negative) kinetic energy of the $\chi$ field grows than that of
the $\varphi$ field. The scale factor smoothly evolves through the
bounce and finally moves towards the critical point $B_+$ in the
expanding phase, where once again the model approaches a power-law
solution with $p=2/\lambda^2$.

We use Eqs.~(\ref{vphiqq}-\ref{sq}) to evolve the scalar perturbations 
in the uniform $\chi$-field gauge through the bounce using $q$ as
our monotonically increasing time variable. This requires initial
conditions for $\delta\vphi,~ \delta\vphi_{,q},~ \psi, ~\psi_{,q},
~\phi$ and $s$ in each mode. The first four initial conditions are
found from our analytical solutions in section(\ref{earlyuc}).
However, as these solutions are found in the uniform curvature
gauge it is necessary to transform them to the uniform
$\chi$-field gauge. It is also necessary to change the time
variable from $\eta$ to $q$. The appropriate gauge transformation
(\ref{xichi}) gives
 \bea
\delta\vphi|_\chi&=&\delta\vphi|_\psi-\alpha\delta\chi|_\psi\,,\\
\delta\vphi_{,q}|_\chi&=& -\frac{\lambda\sqrt6 \beta^2}{2}\delta\chi|_\psi+\frac{\sqrt6}{\kappa\chi'}(\delta\vphi'|_\psi-\alpha\delta\chi'|_\psi)\,,\\
\psi|_\chi&=&\frac{\kappa\gamma}{\sqrt6}\delta\chi|_\psi \,,\\
\psi_{,q}|_\chi&=& \frac{3\kappa\beta^2}{\sqrt6}\delta\chi|_\psi+\frac{\gamma}{\chi'}\delta\chi'|_\psi  \,.
 \eea
The other two initial conditions required, for $\phi$ and $s$, are
found from the constraints (\ref{con1},\ref{con2}).
Thereafter the constraint equations serve as a consistency check
of our numerical solutions.

\begin{figure}
\begin{center}
\includegraphics[width=100mm]{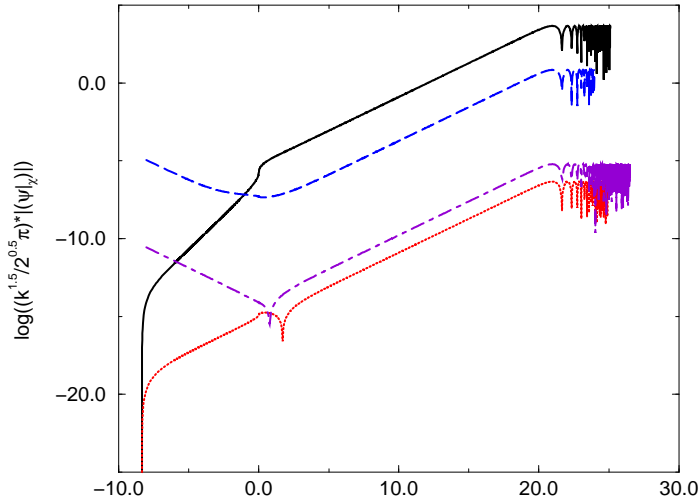}
\caption[psirad]{\label{psirad} Evolution of $\psi|_\chi$ shown as a
  function of expansion, $N=\pm\ln(a/a_0)$, for
  bounce solution with $\lambda=2$ model for each of the four
  independent modes: $\varphi_Y$ (black solid line), $\varphi_J$ (red 
  dotted  line), $\chi_Y$ (blue dashed line), and $\chi_J$ (purple 
  dot-dashed line).} 
\end{center}
\end{figure}

\begin{figure}
\begin{center}
\includegraphics[width=100mm]{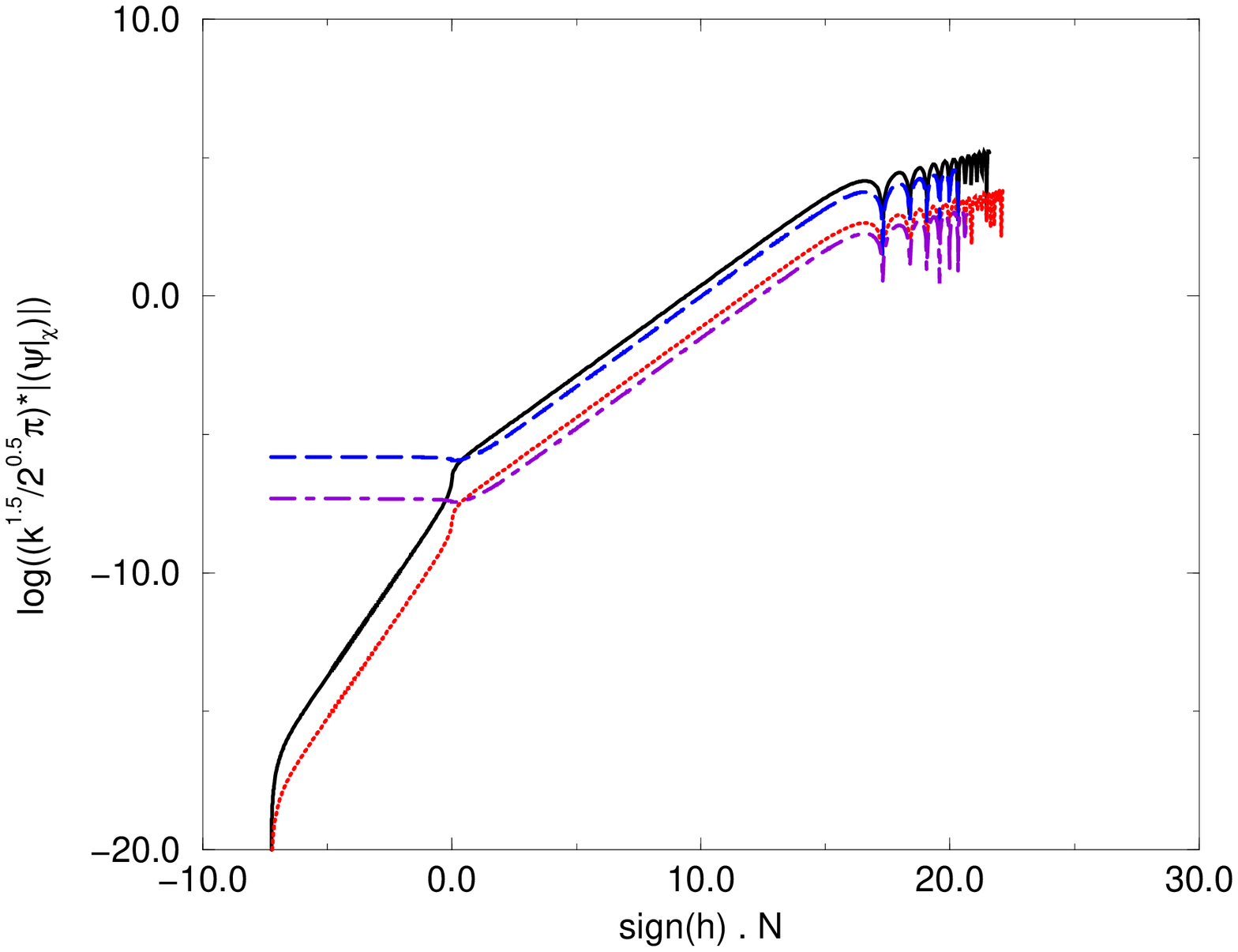}
\caption[psimat]{\label{psimat} Evolution of $\psi|_\chi$ shown as a
  function of expansion, $N=\pm\ln(a/a_0)$, for
  bounce solution with $\lambda=\sqrt3$ model for each of the four
  independent modes: $\varphi_Y$ (black solid line), $\varphi_J$ (red dotted
  line), $\chi_Y$ (blue dashed line), and $\chi_J$ (purple dot-dashed line).}
\end{center}
\end{figure}

The dimensionless metric perturbations $\psi$ and $\phi$ remain
small in the uniform $\chi$-field gauge and well-behaved through
the bounce. Typical numerical solutions for the curvature
perturbation $\psi$ is shown in Figure \ref{psirad} for
$\lambda=2$ and Figure \ref{psimat} for $\lambda=\sqrt3$. The
perturbed lapse function $\phi$ closely follows the curvature
perturbation on large scales, where we find
$\phi+3\psi\simeq$constant, as expected from Eq.~(\ref{phiq}).

The shear potential $\sigma$ does grow rapidly in the collapse
phase, reflecting the instability with respect to shear of the
power-law solution. {}From Eq.~(\ref{rel3}) we see that in any
gauge where $\phi$ and $\psi$ remain small, we find
$\sigma\propto a^{-2}$ as the collapse proceeds. 
However the shear only enters the evolution equation (\ref{phiq}) for
$\phi$ through its spatial divergence and this term remains small as
the bounce is approached. We have checked that the magnitude of the
shear term in, for example, the energy constraint equation
(\ref{con2}), is always 
much smaller than, for instance, the energy density of either of the
scalar fields. Note that the total density, and hence the isotropic
expansion rate, goes to zero at the bounce, so the anisotropic shear
perturbation cannot be smaller than the isotropic expansion/collapse
at this point but it is always small relative to the largest terms in
the equation so we believe our perturbative analysis is justified.

All the $k$-dependent terms in the evolution equations
(\ref{vphiqq}--\ref{sq}) in the
uniform-$\chi$ gauge become small as we approach the bounce and we
find that the $k=0$ solution is an excellent approximation through the
bounce for modes which leave the Hubble scale well before the
bounce ($k\ll (aH)_{\rm max}$). Note the difference with models of
a classical bounce in closed FRW geometries, where finite scale effects 
may be significant as modes re-enter the horizon during the 
bounce~\cite{MP}.

At late times, in the expanding phase after the bounce, the
curvature perturbation in the uniform $\chi$-field gauge gradually
grows and eventually may become larger than one. This is because
$\chi'\to0$ as we asymptotically approach the power-law solution
once again and the gauge shift (\ref{xichi}) becomes large. Thus
at late times in the expanding universe we should transform back
to some well-behaved gauge, such as the longitudinal or
uniform-curvature gauges, where the metric perturbations remain
small. Nonetheless the uniform $\chi$-field gauge does the job of
evolving the initial vacuum fluctuations from the collapse phase
smoothly through the bounce into an expanding phase where the
large-scale behaviour is unambiguous.

\begin{figure}
\begin{center}
\includegraphics[width=70mm]{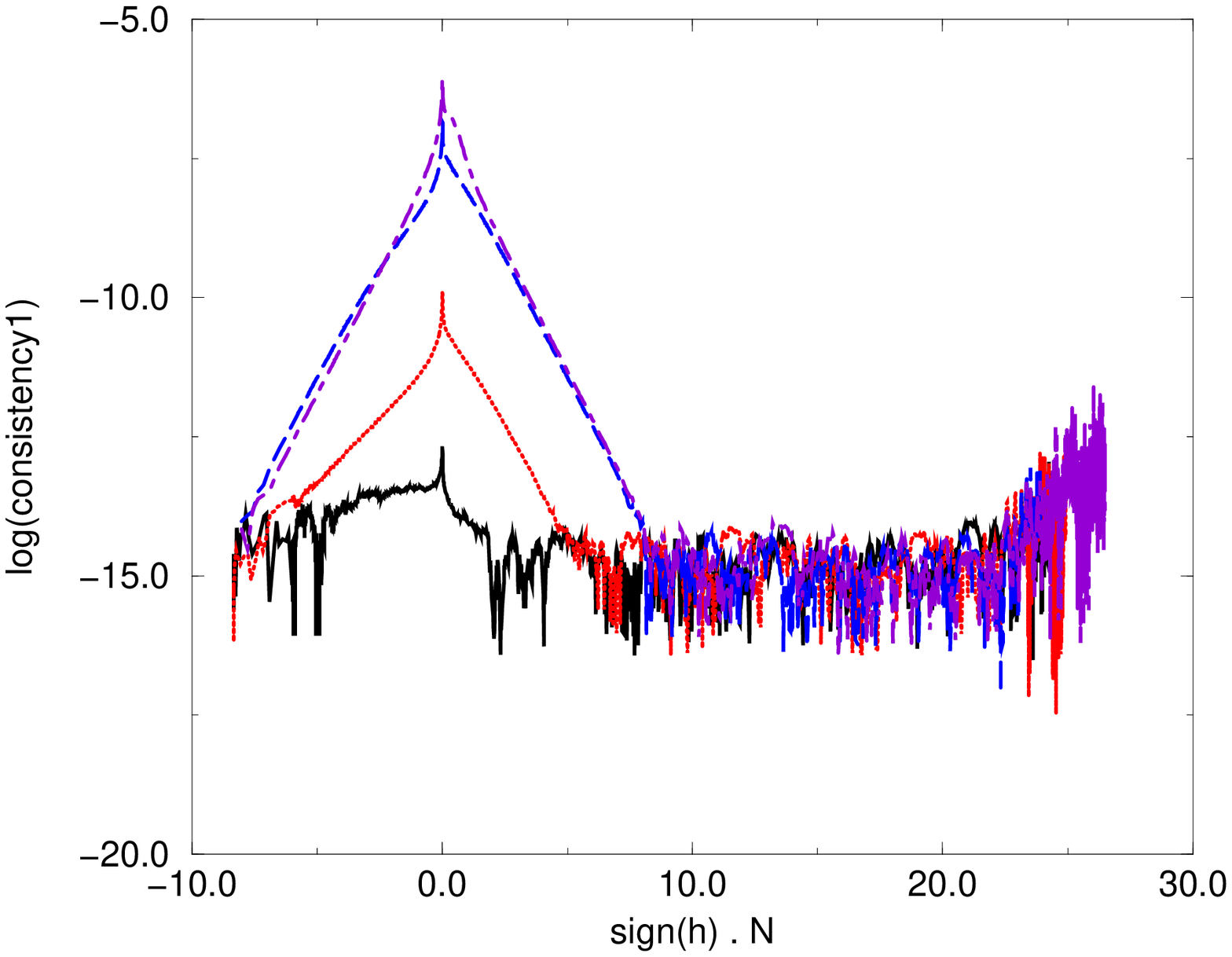}
\includegraphics[width=70mm]{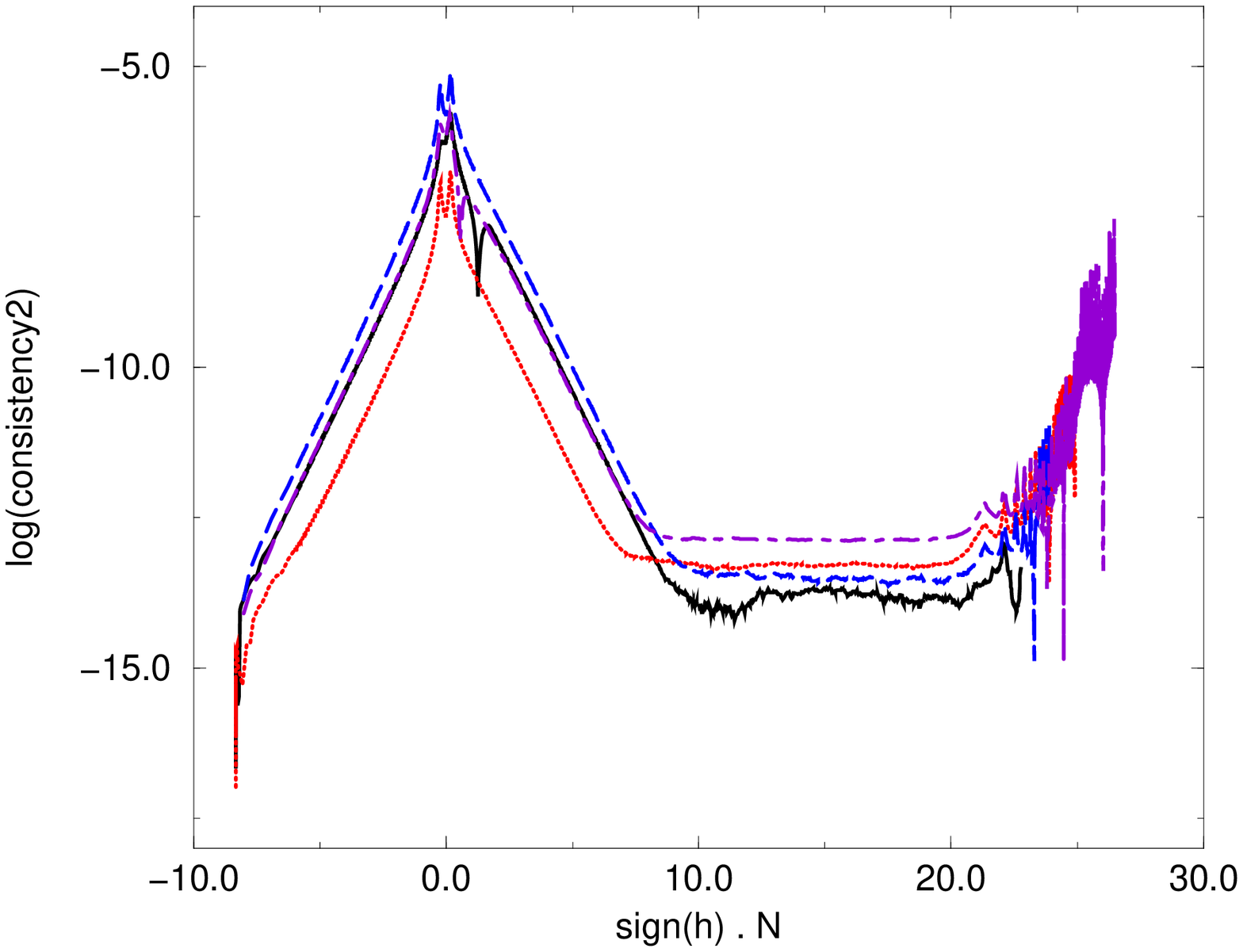}
\end{center}
\caption[consistlambda2]{\label{consistlambda2} The violation of constraint 
equations (\ref{con1}) and (\ref{con2}) shown as a function of expansion, 
$N=\pm\ln(a)$, for a bounce solution in the $\lambda=2$ model for each of the 
four independent scalar modes: $\vphi_Y$ (black solid line), 
$\vphi_J$ (red dotted line), $\chi_Y$ (blue dashed line), 
$\chi_J$ (purple dot-dashed line).
The $y$-axis  of the left hand graph shows the logarithm of the 
absolute value of the left hand side of Eq.~(\ref{con1}) divided 
by the sum of the absolute values of each term in the equation. 
The $y$-axis of the right hand graph shows the same thing for 
Eq.~(\ref{con2}).}
\end{figure}

\begin{figure}
\begin{center}
\includegraphics[width=70mm]{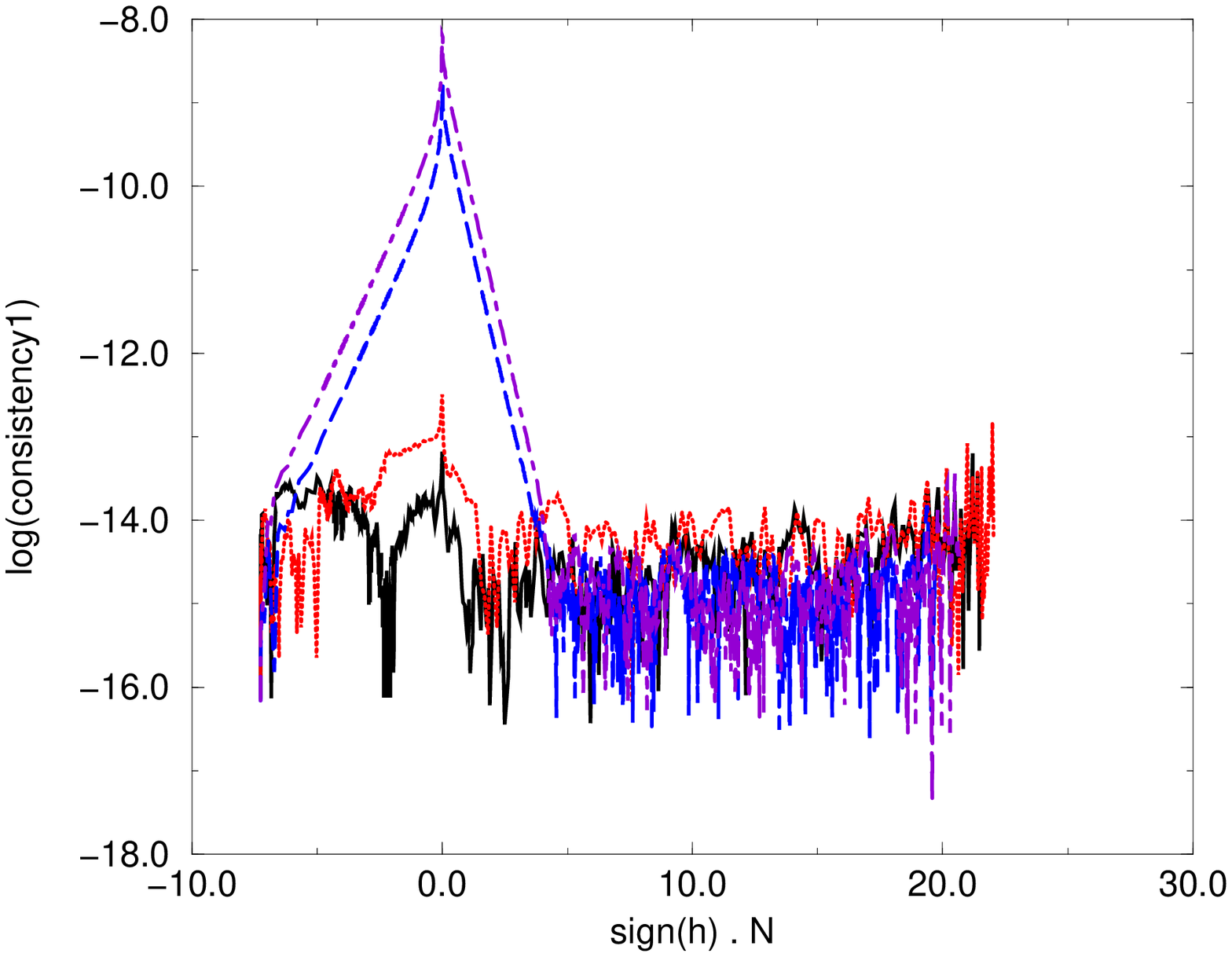}
\includegraphics[width=70mm]{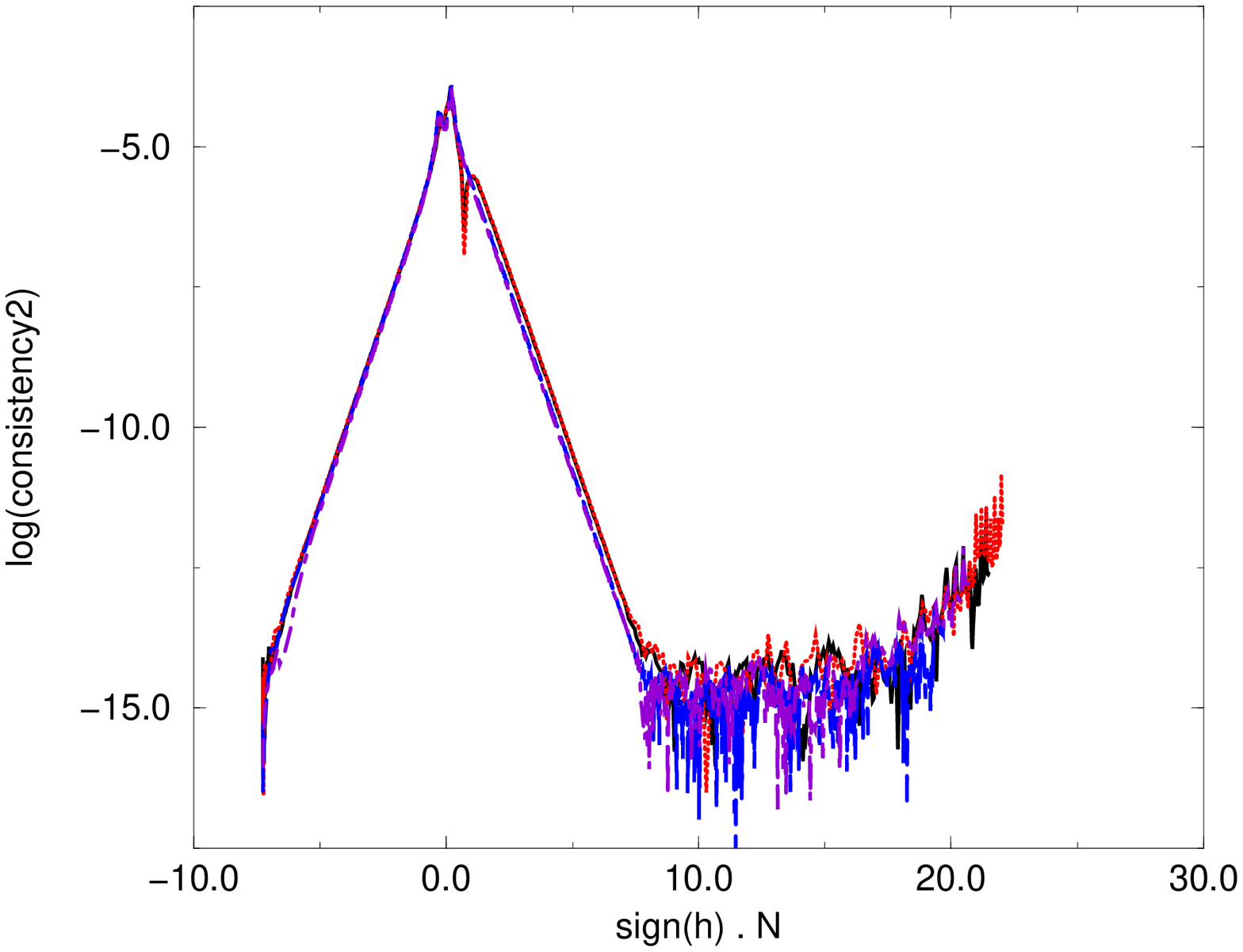}
\end{center}
\caption[consistlambdasqrt3]{\label{consistlambdasqrt3} 
The violation of constraint 
equations (\ref{con1}) and (\ref{con2}) shown as a function of expansion, 
$N=\pm\ln(a)$, for a bounce solution in the $\lambda=\sqrt3$ model for 
each of the 
four independent scalar modes: $\vphi_Y$ (black solid line), 
$\vphi_J$ (red dotted line), $\chi_Y$ (blue dashed line), 
$\chi_J$ (purple dot-dashed line).
The $y$-axis  of the left hand graph shows the logarithm of the 
absolute value of the left hand side of Eq.~(\ref{con1}) divided 
by the sum of the absolute values of each term in the equation. 
The $y$-axis of the right hand graph shows the same thing for 
Eq.~(\ref{con2}).}
\end{figure}

The constraint equations (\ref{con1}) and~(\ref{con2}) are satisfied
to one part in ten thousand or better for all our numerical runs 
(see Figures \ref{consistlambda2} and \ref{consistlambdasqrt3}).

\section{Results for gauge-invariant variables}
\label{results}

In order to compare our results with previous discussions of the
expected evolution of perturbations through a cosmological bounce,
we reconstruct the gauge-invariant variables $\cal{R}$ and $\Phi$
defined in Eqs.~(\ref{defR}) and~(\ref{defPhi}) from the variables
evolved numerically in the uniform-$\chi$ gauge. 

For our two chosen values of $\lambda=2$ and $\lambda=\sqrt{3}$,
we find the power spectra of these two gauge-invariant variables
for each mode when the background is asymptotically evolving as
the power law collapse and expansion. We compare the spectral
tilts of the power spectra before and after the bounce and we see
whether the mode dominant before the bounce remains so after it.

Tensor perturbations $\delta g$ are automatically gauge-invariant and
provide an interesting comparison with the scalar metric
perturbations. Tensors also turn out to place an important constraint
on models producing a scale-invariant spectrum of large-scale
perturbations.

\subsection{Comoving curvature perturbation}

\begin{figure}
\begin{center}
\includegraphics[width=100mm]{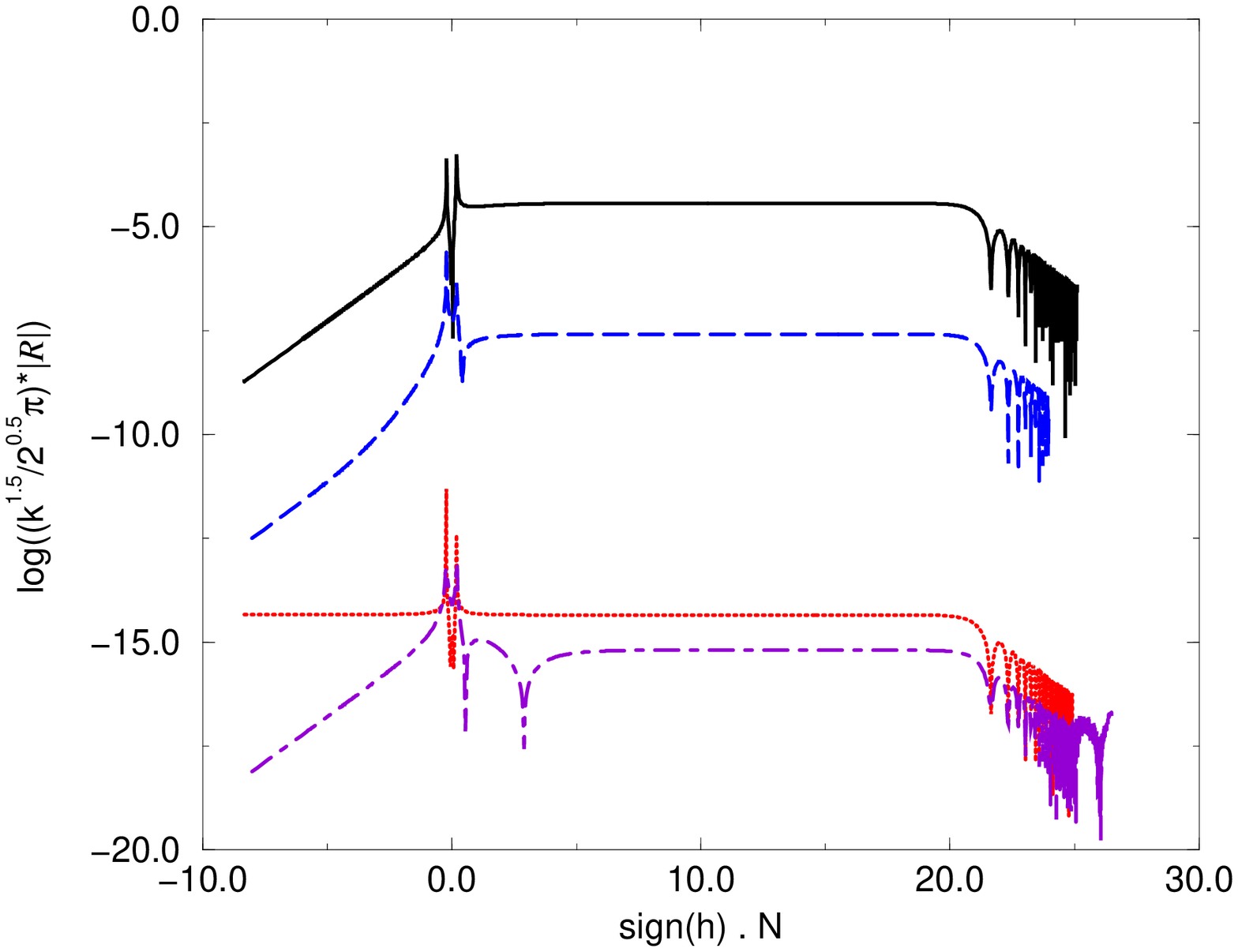}
\caption[Rrad]{\label{Rrad} Comoving curvature perturbation $\cal{R}$
  calculated from numerical solution for each of the four modes shown
  in Figure~\ref{psirad}.} 
\end{center}
\end{figure}

\begin{figure}
\begin{center}
\includegraphics[width=100mm]{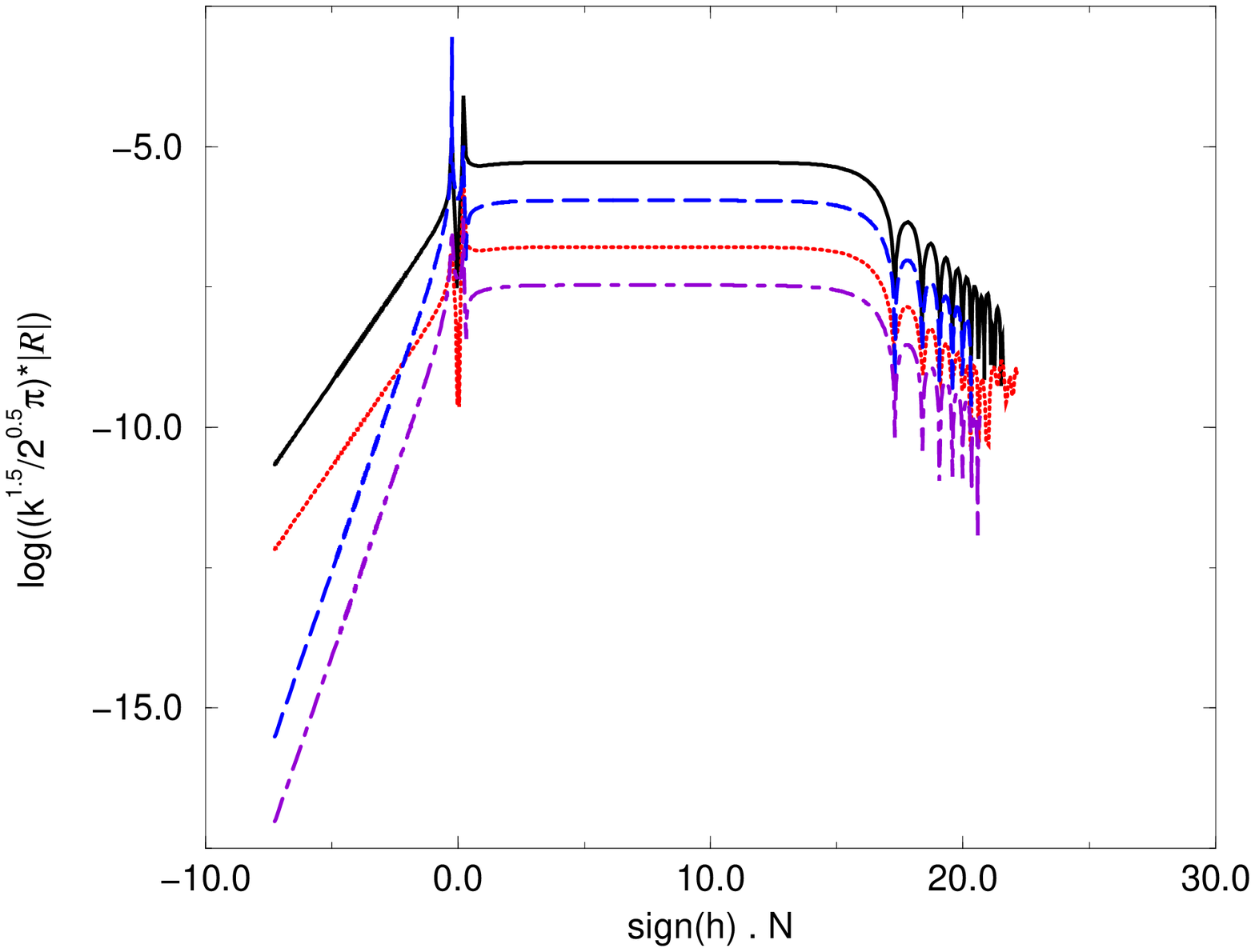}
\caption[Rmat]{\label{Rmat} Comoving curvature perturbation $\cal{R}$
  calculated from numerical solution for each of the four modes shown
  in Figure~\ref{psimat}.}
\end{center}
\end{figure}

Typical evolution of the comoving curvature perturbation, ${\cal R}$
[defined in Eq.~(\ref{defR})], through the bounce is shown in
Figures~\ref{Rrad} for $\lambda=2$ and~\ref{Rmat} for
$\lambda=\sqrt{3}$. The four lines correspond to the four modes
identified in section~\ref{earlyuc}. ${\cal R}$ only remains constant
on large scales for the $\varphi_J$ mode when $\lambda=2$, when this
corresponds to an adiabatic perturbation of the power-law solution.
All the other modes represent non-adiabatic perturbations about the
initial power-law solution on large scales and have non-constant
${\cal R}$. In particular the dominant mode, $\varphi_Y$, describes
non-adiabatic perturbations of the $\varphi$ field. Perturbations in
the $\chi$ field are always sub-dominant in terms of their contribution
to the comoving curvature perturbation.

During the bounce phase we see two singular points in the
evolution of ${\cal R}$. This occurs whenever the total kinetic
energy vanishes, $\varphi^{\prime2}-\chi^{\prime2}=0$
(corresponding to $\alpha^2=1$ in the phase-plane), as can be seen
from the definition of ${\cal R}$ given in Eq.~(\ref{defR}). At
these points the total four-velocity of the two-field system is
momentarily ill-defined. This makes the comoving gauge unsuitable
for evolving the perturbations close to the bounce. We emphasise
though that we can nonetheless trust the evolution of ${\cal R}$
shown in Figures~(\ref{Rrad}) and~(\ref{Rmat}) because here the
comoving curvature has been constructed from perturbations in the
uniform-$\chi$ gauge which remain well-behaved throughout the
bounce, i.e., we do not require that the perturbations are
necessarily small or well-defined in any other gauge.

After the bounce the comoving curvature perturbation rapidly
settles down to a constant value for all modes on all super-Hubble
scales, representing the resultant adiabatic curvature
perturbation on large scales. Thus all the modes contribute a
finite amplitude to the large-scale curvature perturbation in the
expanding phase, and this is approximately equal to the amplitude
of the comoving curvature perturbation at the end of the collapse
phase (but not in general the amplitude at Hubble-crossing during
collapse). The dominant mode at late times is the $\varphi_Y$ mode
that also dominates the comoving curvature perturbation in the
collapse phase.

Finally the perturbations ``re-enter the horizon'' in the
expanding universe and oscillate for $k>aH$.

\subsection{Bardeen potential}

\begin{figure}
\begin{center}
\includegraphics[width=100mm]{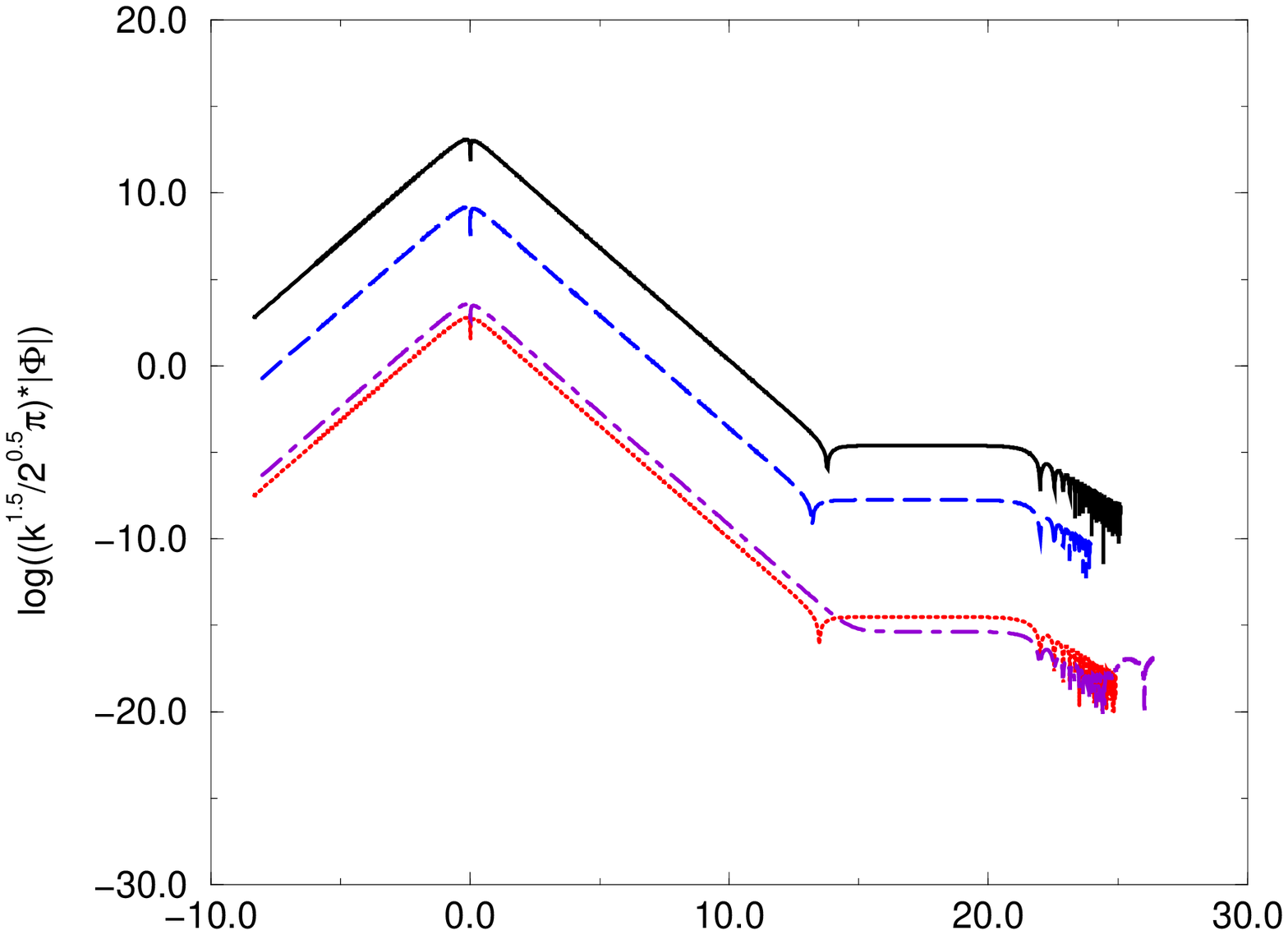}
\caption[Phirad]{\label{Phirad} Bardeen metric potential $\Phi$ 
  calculated from numerical solution for each of the four modes shown
  in Figure~\ref{psirad}.}
\end{center}
\end{figure}

\begin{figure}
\begin{center}
\includegraphics[width=100mm]{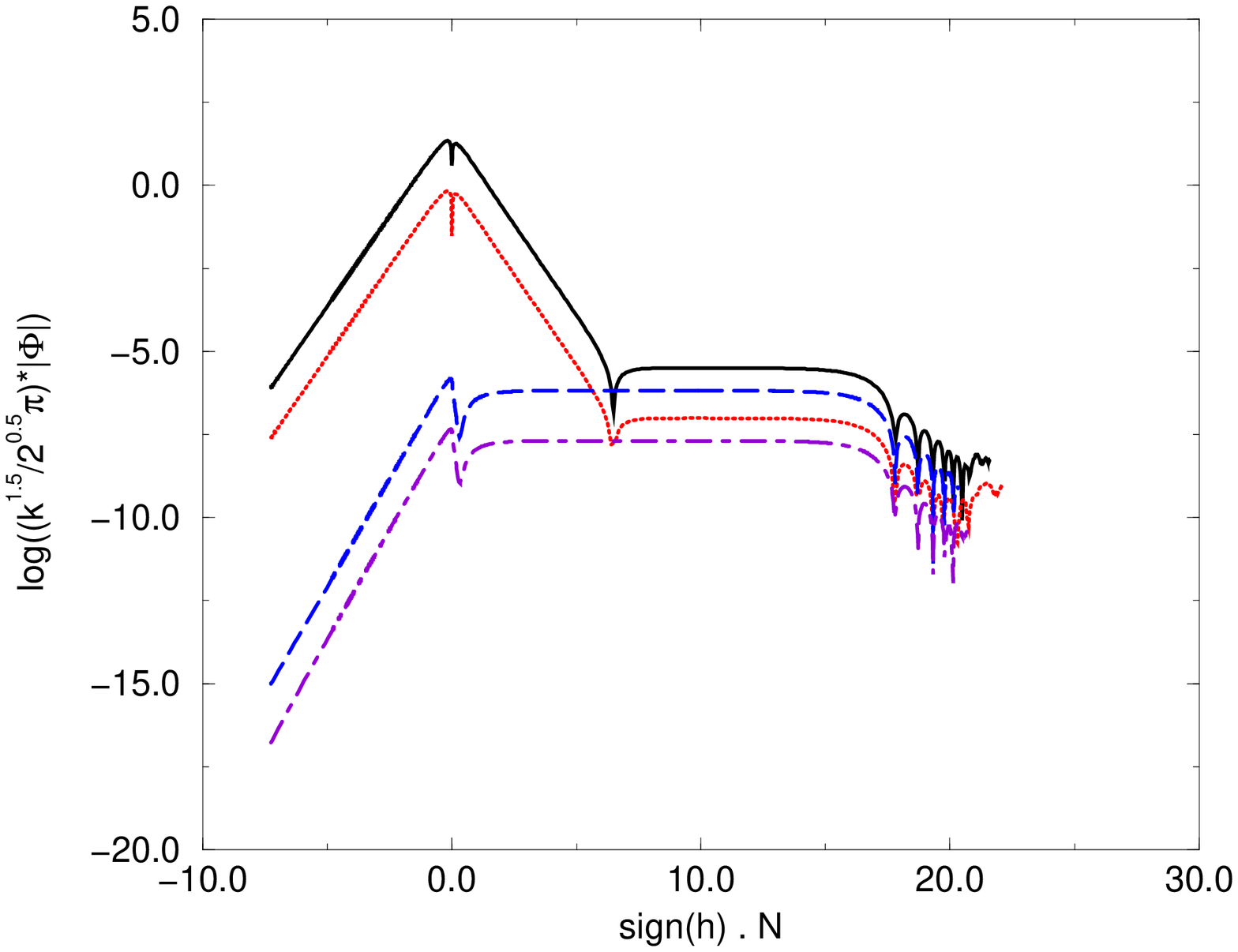}
\caption[Phimat]{\label{Phimat} Bardeen metric potential $\Phi$
  calculated from numerical solution for each of the four modes shown
  in Figure~\ref{psimat}.}
\end{center}
\end{figure}

The typical evolution of the Bardeen potential, $\Phi$ defined in
Eq.~(\ref{defPhi}), which describes the curvature perturbation in
the longitudinal gauge, is shown in Figures~\ref{Phirad} for
$\lambda=2$ and~\ref{Phimat} for $\lambda=\sqrt{3}$.

All modes show the same rapid growth of $\Phi$ during the collapse
phase. We can interpret this as being due to the rapid growth of
the shear in the uniform-$\chi$ gauge, $\sigma|_\chi\propto
a^{-2}$, while the curvature perturbation $\psi|_\chi$ remains
small. According to the definition of $\Phi$ in Eq.~(\ref{defPhi})
this yields a large curvature perturbation in the longitudinal
gauge
 \be
  \Phi \propto \frac{h}{a^2} \,.
 \ee
Our numerical evolution for $\lambda=2$ and $\lambda=\sqrt{3}$
during the power-law collapse phase confirms this, yielding
$\Phi\propto |\eta|^{-3}$ and $\Phi\propto |\eta|^{-5}$ respectively.

Thus the curvature perturbation in the longitudinal gauge becomes
large in a collapse phase meaning that this gauge too may not be
reliable gauge for performing a perturbative calculation through a
cosmological bounce. Once again though we emphasise that our
calculations of the Bardeen potential are reliable as we
reconstruct the Bardeen potential from perturbative calculations
in the uniform-$\chi$ gauge which remain small through the bounce.

After the bounce the dominant modes contributing to the Bardeen
potential, $\varphi_Y$ and $\chi_Y$, remain dominated by the
shear, $\sigma|_\chi$, which now decays in the expanding phase.
Eventually the Bardeen potential settles down to a constant value
on super-Hubble scales, before finally oscillating after
re-entering the Hubble scale. For $\lambda=\sqrt{3}$ the
sub-dominant modes, $\varphi_J$ and $\chi_J$, have much smaller
shear (indeed $\Phi$ stays small for these modes right up until
the bounce) and these modes rapidly settle down to a constant
value soon after the bounce, but this final amplitude is smaller than
that due to the dominant $\varphi_Y$ and $\chi_Y$ modes.

In summary we find that the amplitudes of the Bardeen potential during
collapse are not a good predictor of the amplitude of the eventual
adiabatic curvature perturbation on large scales after the bounce,
even though it becomes much larger than the comoving curvature
perturbation during the collapse. It is the growth of the shear that
dominates the behaviour of $\Phi$ during the collapse and this shear
then decays in the expanding phase. Only once the shear has decayed,
sometime after the bounce, does $\Phi$ settle down to a constant value
on super-Hubble scales in the expanding era.

For the case of $\lambda=2$ corresponding to a radiation-like
equation of state $w=1/3$, we find that once the Bardeen potential
has settled down to a constant value on super-Hubble scales, we
have $\Phi=(2/3){\cal R}$. Similarly for $\lambda=\sqrt{3}$,
corresponding to an equation of state $w=0$, we find
$\Phi=(3/5){\cal R}$, as expected for the growing mode of
adiabatic density perturbations on large scales.

\subsection{Spectral tilts}

\subsubsection{$\lambda=2$: Asymptotic power-law solution with $p=1/2$}

Table~\ref{radtilts} shows the initial spectral tilts of
the power spectra of $\cal{R}$ and $\Phi$, during power-law
collapse with $\lambda=2$. All the modes of $\cal R$ have a blue
spectrum, but the dominant mode of $\Phi$ has a red spectrum.
Table~\ref{radtilts} also shows the spectral tilts of the
super-Hubble power spectra of $\cal{R}$ and $\Phi$ when they have
settled to constant values after the bounce. The tilts of the
spectra of $\cal R$ for each mode remain the same after the bounce
as they were initially. The spectral tilt of $\Phi$ for every mode
however becomes the same as that for the corresponding $\cal R$
spectra
 \be
\Delta n_{\Phi out}=\Delta n_{\Phi in}+4=\Delta n_{\cal R} \,.
 \ee
 
Our model with $\lambda=2$ is very similar to the bounce model
 studied in Ref.~\cite{PPN}, but our results are strikingly different.
 The authors in that paper considered a radiation fluid which shows
 the same initial red spectrum for $\Phi$. But they found that the
 final spectrum for $\Phi$ retained the same red spectrum for
 $\Phi$. There are subtle differences between a radiation fluid and a
 $\lambda=2$ scalar field. In particular the bounce must be
 symmetrical with perfect fluids, whereas our bounce is only
 symmetrical for $C=0$ and our scalar field supports non-adiabatic
 pressure perturbations. But the main difference may come from the
 fact that a second-order evolution equation for $\Phi$ was used to
 follow the evolution of the metric perturbations in
 Ref.~\cite{PPN}. Perturbations in $\Phi$ become large and we have
 instead followed the evolution of perturbations in a gauge which
 remains well-behaved and then reconstructed the value of
 $\Phi$. 
 We have not used the constraint equations to eliminate
 variables and have instead kept used these as a consistency check of
 our numerical integration.

\begin{table}
\begin{center}
\begin{tabular}{c|c|c|c|c}
mode~ & ~$\Delta n_{\cal{R}_{\rm in}}$~ & ~$\Delta n_{\Phi_{\rm in}}$~ & ~
 $\Delta n_{\cal{R}_{\rm out}}$~ & 
 $\Delta n_{\Phi_{\rm out}}$ \\ \hline
$\varphi_Y$~ & 2 & -2 & 2 & 2 \\
$\varphi_J$~ & 4 &  0 & 4 & 4 \\
$\chi_Y$~& 2 & -2 & 2 & 2 \\
$\chi_J$~& 4 &  0 & 4 & 4 \\
\end{tabular}
\caption{Initial spectral tilts for different modes during collapse
 and final tilts after bounce for model with $\lambda=2$.}
\label{radtilts}
\end{center}
\end{table}

\subsubsection{$\lambda=\sqrt3$: Asymptotic power-law solution with $p=2/3$}

Table~\ref{mattilts} shows the initial spectral tilts of
the super-horizon power spectra of $\cal{R}$ and $\Phi$, during
power-law collapse. The dominant mode of $\cal R$ has a
scale-invariant spectrum~\cite{Wands99,FB}, whereas the
spectrum of the dominant mode of $\Phi$ is red.
Table~\ref{mattilts} also shows the spectral tilts of $\cal{R}$
and $\Phi$ when they have settled to constant values on
super-Hubble scales after the bounce. As for $\lambda=2$, the
tilts of the spectra of $\cal R$ for each mode remain the same
after the bounce as they were initially. In particular the
dominant $\cal R$ mode retains its scale-invariant spectrum. The
spectra of $\Phi$ again become the same as ${\cal R}$ after the
bounce on large scales
 \be
\Delta n_{\Phi out}=\Delta n_{\Phi in}+4=\Delta n_{\cal R} \,.
 \ee

Thus we have shown that a power-law collapse phase with $p=2/3$ can
yield a scale-invariant spectrum of curvature perturbations on
super-Hubble scales in the subsequent expanding phase.

\begin{table}
\begin{center}
\begin{tabular}{c|c|c|c|c}
mode~ & ~$\Delta n_{\cal{R}_{\rm in}}$~ & ~$\Delta n_{\Phi_{\rm in}}$~ & ~
 $\Delta n_{\cal{R}_{\rm out}}$~ & 
 $\Delta n_{\Phi_{\rm out}}$ \\ \hline
$\varphi_Y$~&~0~&~-4& 0 & 0\\
$\varphi_J$~&~2~&~-2& 2 & 2\\
$\chi_Y$~&~0~&~-4& 0 & 0\\
$\chi_J$~&~2~&~-2& 2 & 2\\
\end{tabular}
\caption{Initial spectral tilts of different modes during collapse and
 final tilts after bounce for model with $\lambda=\sqrt3$.}
\label{mattilts}
\end{center}
\end{table}

\subsection{Tensor perturbations}

\begin{figure}
\begin{center}
\includegraphics[width=100mm]{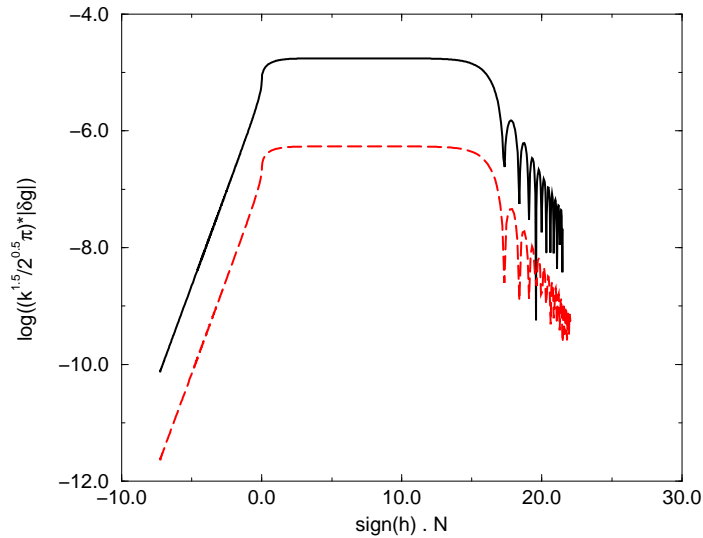}
\caption[deltagmat]{\label{deltagmat} Evolution of $\delta g$ shown as a
  function of expansion, $N=\pm\ln(a)$, for a
  bounce solution in the $\lambda=\sqrt3$ model for each of the two 
  independent tensor modes: $g_Y$ (solid black line) and $g_J$ (dashed
  red line).} 
\end{center}
\end{figure}

The smooth evolution through the bounce of the amplitude of the tensor
metric perturbations is shown in Figure~\ref{deltagmat}. Tensor
perturbations grow rapidly during the collapse phase but then rapidly
settle down to a constant value after the bounce.

The spectral index of the tensor perturbations is the same as that of
the comoving curvature perturbation ${\cal R}$ during collapse and,
like the comoving curvature perturbation, the spectral index is
unaffected by the bounce.

During power-law collapse there is a simple relation between the
amplitude of tensor perturbations and the comoving curvature
perturbation:
\begin{equation}
\left( \frac{{\cal P}_g}{{\cal P}_{\cal R}} \right)_{\rm collapse} 
= 8\kappa^2 \left(
  \frac{\vphi'}{h}\right)^2  \,.
\end{equation}
For collapse solutions described by point $B$, we have, from
Eqs.~(\ref{pointB}) and~(\ref{powerlaw}), 
\begin{equation}
\left( \frac{{\cal P}_g}{{\cal P}_{\cal R}} \right)_{\rm collapse} 
 = \frac{16}{p} \,.
\label{tsratiocollapse}
\end{equation}
In the model with $\lambda=\sqrt3$ which yields a scale-invariant
spectrum of both scalar and tensor perturbations, this tensor-scalar
ratio is $24$ during collapse.

\begin{figure}
\begin{center}
\includegraphics[width=100mm]{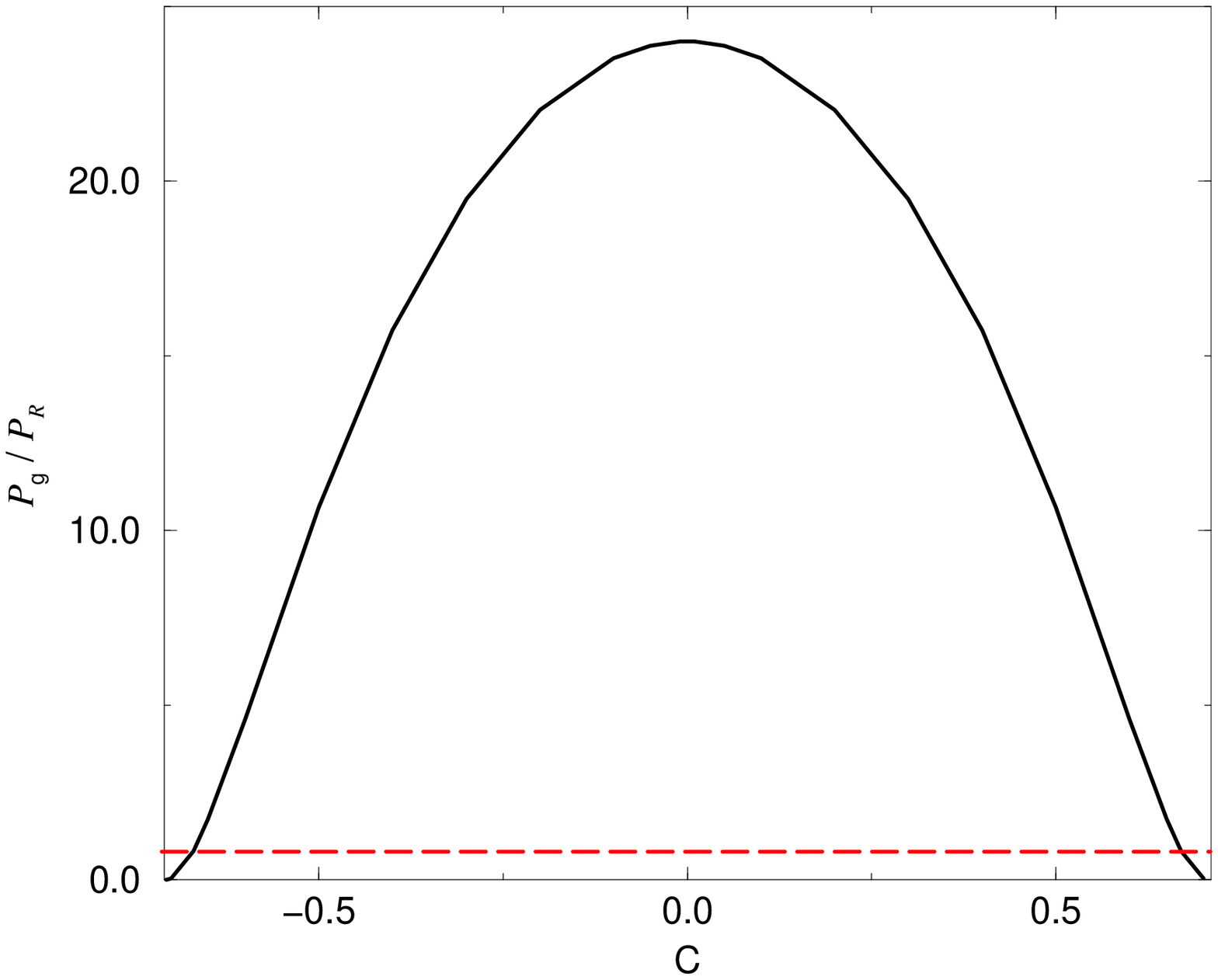}
\caption[tensorscalar]{\label{tensorscalar} Tensor-scalar ratio ${\cal
    P}_g/{\cal P}_{\cal R}$ after the bounce shown against
    dimensionless parameter $C$ defined in Eq.~(\ref{defC}).} The red dashed 
line shows the observational limit. 
\end{center}
\end{figure}

However both the comoving curvature perturbation and tensor
perturbation receive a scale-independent amplification during
the bounce. This amplification is dependent upon the details
of the bounce, and in particular the parameter $C$ defined in
Eq.~(\ref{defC}) which identifies the phase-space trajectory. We show
in Figure~\ref{tensorscalar} the resulting tensor-scalar ratio after
the bounce. The tensor-scalar ratio is maximised for a symmetric
bounce, $C=0$, which leaves the tensor-scalar ratio (\ref{tsratiocollapse}) 
unchanged. For an asymmetric bounce, the scalars are more strongly 
amplified than the tensor modes and the tesnor-scalar ratio becomes 
small for the most asymmetric bounces.

Observations require that the primordial tensor-scalar ratio is less
than 0.81 \cite{WMAP}. We see that this places a severe constraint on
the allowed parameter $|C|>0.67$, given that from Eq.~(\ref{Climit}) 
we require $|C|<0.71$ for the collapse phase to end in a 
non-singular bounce.
Although we have a tight constraint in our particular bounce model, 
it is not clear how problematic this might be in the more general case 
as asymmetry through the bounce can decrease the tensor-scalar ratio.

\section{Conclusions}
\label{conc}

We have presented a simple model of a four-dimensional FRW universe
that describes non-singular bouncing solutions. To do so within the
context general relativity it is well-known that the null-energy
condition must be violated. To achieve this we have introduced a
massless ``ghost'' field with negative kinetic energy. A similar role
is played by higher-order corrections to the low-energy string
effective action which have recently been studied by
Cartier~\cite{Cartier} who found qualitatively similar results. In
order to keep the dynamics as transparent as possible we have
restricted ourselves to a canonical (second-order) scalar field
Lagrangian albeit with the wrong sign. Like the higher-order loop
corrections, our ghost field only plays a significant dynamical role
in the high energy regime close to the bounce.

We have performed a phase-plane analysis of the qualitative
behaviour of expanding and contracting FRW models with both a
canonical scalar field (with positive energy density) and an
exponential potential plus the massless ghost field. Away from the
bounce we have usual power-law expansion or contraction for a
universe dominated by a scalar field with exponential potential
\cite{LM85,Halliwell,BB}. The dimensionless slope of the
potential, $\lambda$, determines the asymptotic power-law index
$p=2/\lambda^2$ and the equation of state $w=(\lambda^2-3)/3$ for
sufficiently flat potentials with $\lambda^2<6$.

It is known that for $\lambda^2<6$ these power-law collapse
solutions describe the generic early time behaviour but are
unstable at late times to kinetic-dominated collapse with $w=1$
\cite{Heard,Gratton} or to shear anisotropy~\cite{Kunze}. Some
kind of instability is a necessary to allow an escape from the
singular power-law collapse. For steep potentials with
$\lambda^2\geq6$ we find no non-singular solutions. But for flat
potentials with $\lambda^2<6$ we find a class of non-singular
solutions that begin in the asymptotic past as a power-law
collapse, approach a minimum value of the scale factor and then
approach a power-law expansion at late times.

This class of non-singular solutions provides a simple bounce
model in which we then study the evolution of linear
perturbations. We assume that these perturbations begin in the
flat-spacetime vacuum state at early times and use this to predict
the amplitude of large scale (super-Hubble) adiabatic density
perturbations after the bounce.

Although both the Bardeen potential $\Phi$ and comoving curvature
perturbation ${\cal R}$ evolve smoothly at the instantaneous
bounce, they both have potential problems. The Bardeen potential
grows rapidly during the collapse phase and is much greater than
unity for the modes of interest, calling into question the
validity of linear perturbation theory in the longitudinal
gauge~\cite{Lyth2,PPN1}. On the other hand the comoving
gauge becomes singular whenever the total kinetic energy of the
scalar fields vanishes~\cite{Kodama}, which occurs shortly
before and after the bounce in our model.

We can however find a gauge in which the metric perturbations
remain small and linear perturbation theory should be valid. In our model
the massless ghost field is monotonic and provides a suitable time
variable, defining a well-defined choice of gauge (the uniform 
$\chi$-field gauge) for perturbations. We note that we solve the 
regularized unconstrained evolution equations for the perturbations, 
using the constraint equations as a check of our numerical integration. 
This avoids potential problems that could arise from using the 
constraint equations to eliminate variables whose coefficients 
vanish at the bounce~\cite{Cartier}. 
We can then reconstruct the usual gauge-invariant
variables from calculations in this gauge.

We find that the comoving curvature perturbation calculated during
the collapse phase is a good estimator of the resulting adiabatic
density perturbation on super-Hubble scales in the expanding
phase. Note that the value at the end of the collapse phase is not
in general the same as the that at Hubble-crossing during collapse
(unlike the usual case in an inflationary expansion). This is
because the dominant mode during collapse is a non-adiabatic field
perturbation. However the comoving curvature perturbation remains
constant on super-Hubble scales after the bounce.

By contrast the Bardeen potential calculated in the collapse phase
is not a good indicator of the subsequent large scale adiabatic
perturbation in the expanding phase. It is dominated by the shear
in the uniform $\chi$-field gauge, which grows rapidly in the
collapse phase and then decays away after the bounce, until the
Bardeen potential settles down to a constant value on super-Hubble
scales, related to the comoving curvature perturbation by the
usual expression $\Phi=3(1+w){\cal R}/(5+3w)$. This basic point
has been made by many authors \cite{Lyth,BF,Hwang,LW03}
and here we demonstrate it explicitly for our simple bounce model.

In particular we find that the spectral tilt of the final
adiabatic curvature perturbation on super-Hubble scales in the
expanding phase is given by the spectral tilt of the initial
comoving curvature perturbations (or equivalently the scalar field
fluctuations in the spatially flat gauge). We therefore have given
an explicit model for a power-law collapse with $\lambda=\sqrt{3}$
which yields a scale-invariant of adiabatic perturbations on
super-Hubble scales in the expanding phase.

The amplitude of the comoving curvature perturbation on the
smallest scale (that is the minimum value of the Hubble scale
during collapse) is set by the energy scale relative to the Planck
scale (where vacuum fluctuations in the metric become of order
unity)
 \be
 {\cal P}_{\cal R} (k=aH_{\rm max})
  \sim \left( \frac{H_{\rm max}}{M_{\rm Pl}} \right)^2 \,.
 \ee
This is very similar to the amplitude of gravitational waves (tensor
perturbations) produced during collapse. The tensor-scalar ratio is an
important constraint on conventional inflation models and turns out to
be a severe constraint on our collapse model when
$\lambda=\sqrt{3}$. 

Pre big bang or collapse models have been criticised for the degree of
fine-tuning required to set up the initial conditions at the start of
any collapse phase~\cite{Turner,Linde,Gratton}.
The dimensionless parameter $C$, defined in Eq.~(\ref{defC}), is a
first integral of the equations of motion which selects the trajectory
in the homogeneous phase-space. Although any value for
$|C|<1/\sqrt{2}$ yields a non-singular bounce for models with
$\lambda=\sqrt{3}$, only trajectories with $|C|>0.67$ give an
acceptably small tensor-scalar ratio. This provides a quantitative
measure of the degree of fine-tuning of initial conditions required to
get an acceptable primordial density perturbation within the class of
models considered in this paper.

\acknowledgments

The authors are grateful to Simon Clark, Ed Copeland, Fabio Finelli, 
Ruth Lazcoz and Alexei Starobinsky for useful comments.
LEA is supported by a Particle Physics and Astronomy Research
Council studentship. DW is supported by the Royal Society.




\begin{thebibliography}{99}

\bibitem{WMAP}
H.~V.~Peiris {\it et al.},
Astrophys.\ J.\ Suppl.\  {\bf 148}, 213 (2003)
[arXiv:astro-ph/0302225].

\bibitem{LLbook}
A.~R.~Liddle and D.~H.~Lyth,
2000, {\em Cosmological inflation and large-scale structure},
Cambridge University Press.

\bibitem{GV}
M.~Gasperini and G.~Veneziano,
Astropart.\ Phys.\  {\bf 1}, 317 (1993)
[arXiv:hep-th/9211021];
Phys.\ Rept.\  {\bf 373}, 1 (2003)
[arXiv:hep-th/0207130].

\bibitem{pbbreviews}
J.~E.~Lidsey, D.~Wands and E.~J.~Copeland,
Phys.\ Rept.\  {\bf 337}, 343 (2000)
[arXiv:hep-th/9909061].

\bibitem{ekpyrotic}
J.~Khoury, B.~A.~Ovrut, P.~J.~Steinhardt and N.~Turok,
Phys.\ Rev.\ D {\bf 64}, 123522 (2001)
[arXiv:hep-th/0103239].

\bibitem{pyro}
R.~Kallosh, L.~Kofman and A.~D.~Linde,
Phys.\ Rev.\ D {\bf 64}, 123523 (2001)
[arXiv:hep-th/0104073].

\bibitem{cyclic}
P.~J.~Steinhardt and N.~Turok,
arXiv:hep-th/0111030;
Phys.\ Rev.\ D {\bf 65}, 126003 (2002)
[arXiv:hep-th/0111098].

\bibitem{BGGMV}
R.~Brustein, M.~Gasperini, M.~Giovannini, V.~F.~Mukhanov and G.~Veneziano,
Phys.\ Rev.\ D {\bf 51}, 6744 (1995)
[arXiv:hep-th/9501066].

\bibitem{Lyth}
D.~H.~Lyth,
Phys.\ Lett.\ B {\bf 524}, 1 (2002)
[arXiv:hep-ph/0106153].

\bibitem{curvaton}
K.~Enqvist and M.~S.~Sloth,
Nucl.\ Phys.\ B {\bf 626}, 395 (2002)
[arXiv:hep-ph/0109214];
D.~H.~Lyth and D.~Wands,
Phys.\ Lett.\ B {\bf 524}, 5 (2002)
[arXiv:hep-ph/0110002];
T.~Moroi and T.~Takahashi,
Phys.\ Lett.\ B {\bf 522}, 215 (2001) [Erratum-ibid.\ B {\bf 539},
303 (2002)] [arXiv:hep-ph/0110096].

\bibitem{Wands99}
D.~Wands,
Phys.\ Rev.\ D {\bf 60}, 023507 (1999)
[arXiv:gr-qc/9809062].

\bibitem{FB}
F.~Finelli and R.~Brandenberger,
Phys.\ Rev.\ D {\bf 65}, 103522 (2002)
[arXiv:hep-th/0112249].

\bibitem{Lyth85}
D.~H.~Lyth,
Phys.\ Rev.\ D {\bf 31}, 1792 (1985).

\bibitem{DM}
N.~Deruelle and V.~F.~Mukhanov,
Phys.\ Rev.\ D {\bf 52}, 5549 (1995)
[arXiv:gr-qc/9503050].

\bibitem{MS}
J.~Martin and D.~J.~Schwarz,
Phys.\ Rev.\ D {\bf 57}, 3302 (1998)
[arXiv:gr-qc/9704049].

\bibitem{WMLL}
D.~Wands, K.~A.~Malik, D.~H.~Lyth and A.~R.~Liddle,
Phys.\ Rev.\ D {\bf 62}, 043527 (2000)
[arXiv:astro-ph/0003278].

\bibitem{LW03}
D.~H.~Lyth and D.~Wands,
Phys.\ Rev.\ D {\bf 68}, 103515 (2003)
[arXiv:astro-ph/0306498].

\bibitem{MP}
J.~Martin and P.~Peter,
Phys.\ Rev.\ D {\bf 68}, 103517 (2003)
[arXiv:hep-th/0307077];
Phys.\ Rev.\ Lett.\  {\bf 92}, 061301 (2004)
[arXiv:astro-ph/0312488];
Phys.\ Rev.\ D {\bf 69}, 107301 (2004)
[arXiv:hep-th/0403173].

\bibitem{BF}
R.~Brandenberger and F.~Finelli,
JHEP {\bf 0111}, 056 (2001)
[arXiv:hep-th/0109004].

\bibitem{Khoury2}
J.~Khoury, B.~A.~Ovrut, P.~J.~Steinhardt and N.~Turok,
Phys.\ Rev.\ D {\bf 66}, 046005 (2002)
[arXiv:hep-th/0109050].

\bibitem{Hwang}
J.~c.~Hwang,
Phys.\ Rev.\ D {\bf 65}, 063514 (2002)
[arXiv:astro-ph/0109045].

\bibitem{Lyth2}
D.~H.~Lyth,
Phys.\ Lett.\ B {\bf 526}, 173 (2002)
[arXiv:hep-ph/0110007].

\bibitem{DV}
R.~Durrer and F.~Vernizzi,
Phys.\ Rev.\ D {\bf 66}, 083503 (2002)
[arXiv:hep-ph/0203275].

\bibitem{HwangNoh}
J.~Hwang and H.~Noh,
Phys.\ Lett.\ B {\bf 545}, 207 (2002)
[arXiv:hep-th/0203193].

\bibitem{CDC}
C.~Cartier, R.~Durrer and E.~J.~Copeland,
Phys.\ Rev.\ D {\bf 67}, 103517 (2003)
[arXiv:hep-th/0301198].

\bibitem{Kodama}
H.~Kodama and T.~Hamazaki,
Prog.\ Theor.\ Phys.\  {\bf 96}, 949 (1996)
[arXiv:gr-qc/9608022].

\bibitem{PPN}
P.~Peter and N.~Pinto-Neto,
Phys.\ Rev.\ D {\bf 66}, 063509 (2002)
[arXiv:hep-th/0203013].

\bibitem{PPNG}
P.~Peter, N.~Pinto-Neto and D.~A.~Gonzalez,
JCAP {\bf 0312}, 003 (2003)
[arXiv:hep-th/0306005].

\bibitem{Finelli:2003mc}
F.~Finelli,
JCAP {\bf 0310}, 011 (2003)
[arXiv:hep-th/0307068].

\bibitem{nonsing}
I.~Antoniadis, J.~Rizos and K.~Tamvakis,
Nucl.\ Phys.\ B {\bf 415}, 497 (1994)
[arXiv:hep-th/9305025];

\bibitem{Ed}
C.~Cartier, E.~J.~Copeland and R.~Madden,
JHEP {\bf 0001}, 035 (2000)
[arXiv:hep-th/9910169].

\bibitem{Schwarz}
J.~Martin, P.~Peter, N.~Pinto Neto and D.~J.~Schwarz,
Phys.\ Rev.\ D {\bf 65}, 123513 (2002)
[arXiv:hep-th/0112128];
Phys.\ Rev.\ D {\bf 67}, 028301 (2003)
[arXiv:hep-th/0204222].

\bibitem{Tolley}
A.~J.~Tolley, N.~Turok and P.~J.~Steinhardt,
Phys.\ Rev.\ D {\bf 69}, 106005 (2004)
[arXiv:hep-th/0306109].


\bibitem{Tsujikawa:2001ad}
S.~Tsujikawa,
Phys.\ Lett.\ B {\bf 526}, 179 (2002)
[arXiv:gr-qc/0110124].

\bibitem{Tsujikawa:2002qc}
S.~Tsujikawa, R.~Brandenberger and F.~Finelli,
Phys.\ Rev.\ D {\bf 66}, 083513 (2002)
[arXiv:hep-th/0207228].


\bibitem{Cartier}
C.~Cartier,
arXiv:hep-th/0401036.

\bibitem{CartierHwang}
C.~Cartier, J.~c.~Hwang and E.~J.~Copeland,
Phys.\ Rev.\ D {\bf 64}, 103504 (2001)
[arXiv:astro-ph/0106197].

\bibitem{Brustein}
R.~Brustein and R.~Madden,
Phys.\ Lett.\ B {\bf 410}, 110 (1997)
[arXiv:hep-th/9702043].


\bibitem{ParkerFulling}
L.~Parker, S.~A.~Fulling,
Phys.\ Rev.\ D {\bf 7}, 2357 (1973)

\bibitem{Starobinsky78}
A.~A.~Starobinsky, 
Sov.\ Astron.\ Lett. {\bf 4}, 82 (1978)


\bibitem{Matzner}
J.~D.~Barrow and R.~A.~Matzner,
Phys.\ Rev.\ D {\bf 21}, 336 (1980).

\bibitem{Hwangbounce}
J.~c.~Hwang and H.~Noh,
Phys.\ Rev.\ D {\bf 65}, 124010 (2002)
[arXiv:astro-ph/0112079].

\bibitem{GordonTurok}
C.~Gordon and N.~Turok,
Phys.\ Rev.\ D {\bf 67}, 123508 (2003)
[arXiv:hep-th/0206138].

\bibitem{Deruelle}
N.~Deruelle,
arXiv:gr-qc/0404126.

\bibitem{DeruelleStreich}
N.~Deruelle, A.~Streich,
arXiv:gr-qc/0405003.

\bibitem{TT}
P.~Teyssandier and P.~Tourrenc, 
J.\ Math.\ Phys.\ {\bf 24}, 2793 (1983).

\bibitem{Wands93}
D.~Wands,
Class.\ Quant.\ Grav.\  {\bf 11}, 269 (1994)
[arXiv:gr-qc/9307034].

\bibitem{ghosts}
S.~M.~Carroll, M.~Hoffman and M.~Trodden,
Phys.\ Rev.\ D {\bf 68}, 023509 (2003)
[arXiv:astro-ph/0301273];
J.~M.~Cline, S.~y.~Jeon and G.~D.~Moore,
arXiv:hep-ph/0311312.

\bibitem{Caldwell}
R.~R.~Caldwell,
Phys.\ Lett.\ B {\bf 545}, 23 (2002)
[arXiv:astro-ph/9908168].

\bibitem{CEW}
E.~J.~Copeland, A.~R.~Liddle and D.~Wands,
Phys.\ Rev.\ D {\bf 57}, 4686 (1998)
[arXiv:gr-qc/9711068].

\bibitem{Heard}
I.~P.~C.~Heard and D.~Wands,
Class.\ Quant.\ Grav.\  {\bf 19}, 5435 (2002)
[arXiv:gr-qc/0206085].

\bibitem{LM85}
F.~Lucchin and S.~Matarrese,
Phys.\ Rev.\ D {\bf 32}, 1316 (1985).

\bibitem{Halliwell}
J.~J.~Halliwell,
Phys.\ Lett.\ B {\bf 185}, 341 (1987).

\bibitem{BB}
A.~B.~Burd and J.~D.~Barrow,
Nucl.\ Phys.\ B {\bf 308}, 929 (1988).

\bibitem{MFB92}
V.~F.~Mukhanov, H.~A.~Feldman and R.~H.~Brandenberger,
Phys.\ Rept.\  {\bf 215}, 203 (1992).

\bibitem{MW1}
K.~A.~Malik and D.~Wands,
arXiv:gr-qc/9804046.

\bibitem{Gordon}
C.~Gordon, D.~Wands, B.~A.~Bassett and R.~Maartens,
Phys.\ Rev.\ D {\bf 63}, 023506 (2001)
[arXiv:astro-ph/0009131].

\bibitem{GBW}
J.~Garcia-Bellido and D.~Wands,
Phys.\ Rev.\ D {\bf 53}, 5437 (1996)
[arXiv:astro-ph/9511029].

\bibitem{LythStewart}
D.~H.~Lyth and E.~D.~Stewart,
Phys.\ Lett.\ B {\bf 274}, 168 (1992).

\bibitem{Sasaki}
M.~Sasaki,
Prog.\ Theor.\ Phys.\  {\bf 76}, 1036 (1986).

\bibitem{hwang}
J.~c.~Hwang,
Astrophys.\ J.\  {\bf 427}, 542 (1994).

\bibitem{Starobinsky}
A.~A.~Starobinsky,
JETP Lett.\  {\bf 30}, 682 (1979)
[Pisma Zh.\ Eksp.\ Teor.\ Fiz.\  {\bf 30}, 719 (1979)].

\bibitem{Gratton}
S.~Gratton, J.~Khoury, P.~J.~Steinhardt and N.~Turok,
Phys.\ Rev.\ D {\bf 69}, 103505 (2004)
[arXiv:astro-ph/0301395].

\bibitem{Kunze}
K.~E.~Kunze and R.~Durrer,
Class.\ Quant.\ Grav.\  {\bf 17}, 2597 (2000)
[arXiv:gr-qc/9912081].

\bibitem{PPN1}
P.~Peter and N.~Pinto-Neto,
Phys.\ Rev.\ D {\bf 65}, 023513 (2002)
[arXiv:gr-qc/0109038].

\bibitem{Turner}
M.~S.~Turner and E.~J.~Weinberg,
Phys.\ Rev.\ D {\bf 56}, 4604 (1997)
[arXiv:hep-th/9705035].

\bibitem{Linde}
N.~Kaloper, A.~D.~Linde and R.~Bousso,
Phys.\ Rev.\ D {\bf 59}, 043508 (1999)
[arXiv:hep-th/9801073].









\end{thebibliography}
\end{document}